\documentclass[a4paper,11pt]{article}
\pdfoutput=1
\usepackage{jheppub} 

\usepackage[T1]{fontenc} 
\usepackage{graphicx,xcolor}
\usepackage{float}
\usepackage{mathtools,braket,blkarray}
\allowdisplaybreaks
\usepackage{amssymb,amsfonts,relsize}
\usepackage[width=0.9\textwidth,footnotesize]{caption}
\usepackage[font=scriptsize]{subcaption}
\hypersetup{pageanchor=false}
\usepackage{booktabs,multirow,makecell}
\usepackage[utf8]{inputenc}
\usepackage{cleveref}

\newcommand{\be}{\begin{equation}}
\newcommand{\ee}{\end{equation}}
\newcommand{\bea}{\begin{eqnarray}}
\newcommand{\eea}{\end{eqnarray}}

\newcommand{\overbar}[1]{\mkern 1.5mu\overline{\mkern-1.5mu#1\mkern-1.5mu}\mkern 1.5mu}

\newcommand*\diff{\mathop{}\!\mathrm{d}}

\author[a,b,c]{Julia Gehrlein}
\author[b,c]{and Mathias Pierre}

\affiliation[a]{Physics Department, Brookhaven National Laboratory, Upton, New York 11973, USA}
\affiliation[b]{Instituto de F\'{i}sica Te\'{o}rica (IFT) UAM-CSIC, Campus de Cantoblanco, 28049 Madrid, Spain}
\affiliation[c]{
Departamento de F\'{i}sica Te\'{o}rica, Universidad Autónoma de Madrid (UAM), Campus de
Cantoblanco, 28049 Madrid, Spain}

\emailAdd{jgehrlein@bnl.gov}
\emailAdd{mathias.pierre@uam.es}

\abstract{ We consider a minimal extension of the Standard Model with a hidden sector charged under a dark local $U(1)'$ gauge group, accounting simultaneously for light neutrino masses and the observed Dark Matter relic abundance. The model contains two copies of right-handed neutrinos which give rise to light neutrino-masses via an extended seesaw mechanism. The presence of a stable Dark-Matter candidate and a massless state naturally arise by requiring the simplest anomaly-free particle content without introducing any extra symmetries. We investigate the phenomenology of the hidden sector considering the $U(1)'$ breaking scale of the order of the electroweak scale. Confronting the thermal history of  this hidden-sector model with existing and future constraints from collider, direct and indirect detection experiments provides various possibilities of probing the model in complementary ways as every particle of the dark sector plays a specific cosmological role.
Across the identified viable parameter space, a large region predicts a sizable contribution to the effective relativistic degrees-of-freedom in the early Universe that allows to alleviate the recently reported tension between late and early measurements of the Hubble constant. 

}

\arxivnumber{1912.06661}

\begin{document}

\preprint{IFT-UAM/CSIC-19-167}

\title{A testable hidden-sector model for Dark Matter and neutrino masses}

\maketitle

\section{Introduction}
\label{sec:intro}

The Standard Model (SM) of particle physics has been tested to a great accuracy and successfully describes most of the experimentally observed phenomena from microscopic to cosmological scales. However, there are several open questions which require a beyond-the-SM explanation. Two of the most prominent indications that the SM is incomplete are  the experimental observation of neutrino oscillations and hence non-vanishing neutrino masses which require the addition of a neutrino
mass term to the SM, and the lack of a Dark Matter (DM) candidate within the Standard Model particle content.

Many new physics constructions attempt at explaining neutrino masses at a high scale, for instance the famous type I seesaw~\cite{Minkowski:1977sc,Ramond:1979py,GellMann:1980vs,Yanagida:1979as,Mohapatra:1979ia,Schechter:1980gr} where the smallness of the neutrino masses is explained via a suppression of the electroweak scale. However the same ratio which suppresses the neutrino mass also suppresses the possible signals of these models and brings the new physics particles out of the reach of current and future experiments. Alternatively, one can also explain the smallness of neutrino masses via  an approximate  symmetry which also allows TeV scale new particles in the spectrum. This idea is used in the inverse seesaw~\cite{Mohapatra:1986bd}, linear seesaw~\cite{Akhmedov:1995vm,Malinsky:2005bi} and Double Seesaw~\cite{Mohapatra:1986bd,Mohapatra:1986aw,Roncadelli:1983ty,Roy:1983be}.\\
Dark matter candidates at the electroweak scale received much attention over the past few years in the context of the Weakly Interacting Massive Particles (WIMP) paradigm~\cite{Bertone:2004pz,Arcadi:2017kky}. However constraints from direct and indirect detection have pushed the simplest models such as Higgs~\cite{Djouadi:2011aa,Djouadi:2012zc,Casas:2017jjg,Arcadi:2019lka}
 or $Z$-portal~\cite{Arcadi:2014lta,Lebedev:2014bba,Kearney:2016rng,Balazs:2017ple,Arcadi:2017kky,Escudero:2016gzx} towards corners of viable parameter space, implying the need to introduce extra states and jeopardize the simplicity of these constructions. Therefore simple alternative solutions to the DM conundrum such as the non-thermal DM freeze-in mechanism have been investigated in the context of very-weakly coupled scenarios~\cite{McDonald:2001vt,Hall:2009bx,Chu:2011be,Bernal:2017kxu} or by considering new high energy scales~\cite{Mambrini:2013iaa,Garny:2017kha,Garcia:2017tuj,Bernal:2018qlk,Bhattacharyya:2018evo,Chowdhury:2018tzw,Chianese:2019epo}. Aside from the theoretical motivations for such scenarios, they remain difficult to be probed by present and future experiments, only few specific cases lead to interesting signatures and detection prospects~\cite{Belanger:2018sti,Curtin:2018mvb,Hambye:2018dpi,No:2019gvl}. Another part of the community have explored \textit{hidden sector} models at the electroweak scale where new-physics particles are not directly charged under the SM gauge group but they only interact via portals with the SM particles. For instance in DM models based on the Strongly Interacting Massive Particles (SIMP) or Elastic Decoupling Relic (ELDER) mechanisms, interactions are sizable within the dark sector and the primordial thermalization between both hidden and SM sectors implies potential signatures making such scenarios quite appealing~\cite{Hochberg:2014dra,Hochberg:2014kqa,Bernal:2015xba,Kuflik:2015isi,Kuflik:2017iqs,Choi:2017zww,Hochberg:2018rjs,Choi:2019zeb,Blennow:2019fhy}. \\
Therefore, it is very tempting to consider a simultaneous hidden-sector as a solution to open problems of the SM such as dark matter and neutrino masses, at the GeV scale (which is also motivated by naturalness arguments~\cite{Casas:2004gh, Vissani:1997ys}), while still suggesting detection possibilities in the near future. Such detection prospects could be achievable for instance if thermalization occurs at some point between the SM and hidden-sector in the early universe, implying a sizable coupling between both sectors.

In this manuscript we follow this approach and introduce a new dark $U(1)'$ symmetry under which only particles from the hidden sector are charged. The hidden sector contains right-handed neutrinos which give mass to the SM neutrinos via an extended seesaw (ESS) mechanism and in the simplest anomaly-free version of the theory, additional chiral fermions are introduced in the hidden sector, one of them is a viable DM candidate and one extra state remains massless and contributes to the effective relativistic degrees-of-freedom in the early Universe.
The DM mass in our model has the same origin as the mass of the right-handed neutrinos, namely they obtain their mass from the coupling to a scalar which spontaneously breaks the dark gauge symmetry giving rise to a massive dark gauge boson $Z'$ in the spectrum around the same scale.
Interactions between the dark sector and the SM are possible via only three portals: the mixing of the dark scalar with the Higgs boson, the mixing of the dark $Z^\prime$ gauge boson with the SM $Z$ boson, and the mixing of the right-handed neutrinos with the SM neutrinos induced by a new Yukawa term.
 Furthermore, we predict all new physics particles below or around the electroweak scale, which makes them potentially observable and a good target for current and future experiments.

In this work the phenomenology of the hidden sector is investigated. We find that the DM candidate with a mass around the GeV scale can account for the observed relic density while at the same time evading  bounds from direct and indirect detection. The mixing of the dark gauge boson with the SM $Z$ plays a crucial role in this model as this interaction leads to the thermalization of the dark sector with the SM. We identify the allowed parameter space for which the dark-sector particles evade current constraints but can be probed with future experiments. Moreover, contribution from the hidden-sector states to the effective number of relativistic species in the early Universe can relax the recently reported Hubble tension~\cite{Wong:2019kwg,Verde:2019ivm}.

The paper is organized as follows: After giving an overview of the model in Sec.~\ref{sec:model}, describing the particle content and the Lagrangian, we derive the expression for the neutrino masses and mixings in Sec.~\ref{sec:neutrino}. In Sec.~\ref{sec:ds} and Sec.~\ref{sec:pheno} we analyse the thermal history and phenomenology of the dark sector including the DM density production and constraints on the model, and finally, in Sec.~\ref{sec:results} we summarise and discuss our results.

\section{The model}
\label{sec:model}
This section serves as an overview of the model. The neutrino sector of this model was first proposed in~\cite{Ballett:2019pyw,Ballett:2019cqp}
where the phenomenology of the three portals to the SM and the constraints on the neutrino masses and mixings have been studied. Furthermore, it has been shown that such a model can explain the observed MiniBooNE excess~\cite{Arguelles:2018mtc}.\footnote{A similar model using a hidden sector and dark neutrinos to explain the MiniBooNE excess has been proposed in \cite{Bertuzzo:2018ftf,Bertuzzo:2018itn}.}

The considered model is an ESS neutrino mass model~\cite{Barry:2011wb, Zhang:2011vh,Dev:2012bd}, an extension of the type I seesaw. It was originally introduced to gives rise to a eV sterile neutrino to explain the observed anomalous  disappearance of electron neutrinos at reactors~\cite{Mention:2011rk, Dentler:2017tkw}, and in experiments using intense radioactive  sources~\cite{Acero:2007su,Giunti:2010zu} and appearance of electron neutrinos from a muon neutrino beam at short baseline experiments like LSND~\cite{Aguilar:2001ty} and MiniBooNE~\cite{AguilarArevalo:2010wv, Aguilar-Arevalo:2012fmn,Aguilar-Arevalo:2018gpe}~\footnote{Notice however that the sterile neutrino explanation for these anomalies is  in strong tension with $\nu_\mu$ disappearance searches~\cite{Gariazzo:2017fdh,Dentler:2018sju}.}.
Its usual particle content consists of three additional right-handed neutrinos and one gauge fermion singlet. To obtain an eV scale singlet the right-handed neutrinos need to have masses around $10^{14}$ GeV, like in the type I seesaw, and
similar to a type I seesaw all phenomenological signatures of the right-handed neutrinos are suppressed.

For a model with phenomenological more accessible sterile neutrinos at a lower scale
we choose to 
 introduce more right-handed neutrinos, namely two copies of three right-handed neutrinos $N_R^i$ and $N_R^{\prime i}$  ($i=1,2,3$), but
 only one copy of the additional right-handed neutrinos $N_R^{\prime i}$ is charged under the new dark $U(1)'$. The SM fermions are singlets under $U(1)'$.
 With this charge assignment the dark $U(1)'$ is however anomalous, the 
 triangle anomalies that do not cancel  involve three $U(1)'$ vertices, as well as one $U(1)'$ vertex and  gravity. In order to cancel these anomalies new chiral fermions charged under $U(1)'$ need to be introduced whose charges satisfy
 \begin{equation}
        \sum_i Q^i_L - \sum_i Q^i_R=0~,~
        \sum_i (Q^i_L)^3 - \sum_i (Q^i_R)^3=0~,
    \end{equation}
where $i$ denotes all fermionic states charged under the new symmetry.
A minimal and phenomenological interesting model\footnote{Alternatively, one can introduce 3 left-handed fermions with charge 1 which complexifies the neutrino sector without providing a DM candidate in the model.} is achieved if two left-handed fermions $\chi_L$ and $\omega_L$, and one right-handed fermion $\chi_R$ are added to the SM content in addition to $N_R$ and $N_R'$ needed for the ESS~\footnote{This field content is similar to the model in Ref.~\cite{DeRomeri:2017oxa} which however uses an inverse seesaw with a gauged $B-L$ symmetry to generate neutrino masses.}.
The singlet nature of $N_R$ already allows for a mass term like $\mu_N/2\overline{N_R^c} N_R$ but in order to generate a mass for
$N_R'$  a  complex scalar $\Phi$ charged under $U(1)'$ needs to be introduced. After the scalar spontaneously breaks the dark symmetry and obtains a vacuum expectation value (vev) it generates a mass term for $N_R'$ as well as for $\chi_L,\chi_R$ and gives rise to a massive $U(1)'$ gauge boson $Z'$. The field content is summarized in Tab.~\ref{tab:charges}.
\begin{table}[t]
    \centering
   \begin{tabular}{ c || c | c | c | c | c | c }
    &  $\Phi$ & $N_R$ &$N_R^\prime$ &$\chi_R$ & $\chi_L$ & $\omega_L$ \\
	\hline
    \hline
    $U(1)^\prime$ &  1 & 0 & 1 & 5 &  4 & 4  \\
    \hline
    Multiplicity & 1 & 3 & 3 & 1 & 1 & 1 \\
    \hline
  \end{tabular}
      \caption{Charge assignment and multiplicity of the new fields considered in this model.}
    \label{tab:charges}
\end{table}
With this field content the
complete Lagrangian can be written as 
\begin{equation}
    \mathcal{L}=\mathcal{L}_\text{SM}+\mathcal{L}_\text{scalar}+\mathcal{L}_\text{gauge}+\mathcal{L}_\text{fermions}+\mathcal{L}_\text{kin}~,
\end{equation}
where the scalar sector can be expressed as
\begin{equation}
    \mathcal{L}_\text{scalar}=|D^\mu \Phi|^2+\mu_\phi^2 |\Phi|^2-\lambda_\phi |\Phi|^4-\lambda_{\phi h} |\Phi|^2 |H|^2~,
    \label{eq:Lscalar}
\end{equation}
with $H$ being the SM Higgs doublet and $\Phi=1/\sqrt{2}(v_\phi + \phi)$ in unitary gauge whose vev is denoted by $\langle \Phi\rangle = v_\phi/\sqrt{2}$ and $\phi$ the real scalar degree of freedom. The covariant derivative is $D_\mu \equiv \partial_\mu - i g_{Z^\prime} Z^\prime_\mu$. The gauge sector can be expressed as a sum of kinetic term and kinetic mixing term:
\begin{equation}
    \mathcal{L}_\text{gauge}=-\dfrac{1}{4}X_{\mu \nu} X^{\mu \nu} -\dfrac{\sin \xi}{2} F^{\mu \nu} X_{\mu \nu}~,
\end{equation}
where $X^{\mu \nu}$ and $F^{\mu \nu}$ are the $Z^\prime$ and SM hypercharge field strength tensors  and $\xi$ is the kinetic mixing parameter. The portal between the dark sector and the SM is ensured via the terms $\mathcal{L}_\text{fermions}$ which can be written as
\begin{equation}
    \mathcal{L}_\text{fermions}=-y_\nu^\alpha \overbar{L_L^\alpha} \widetilde{H} N_R - \dfrac{\mu_N}{2} \overbar{N_R^c} N_R - y_N \Phi^* \overbar{N^c_R} N_R^\prime - y_\chi \Phi \overbar{\chi}_R \chi_L+ \text{h.c.}~,
\end{equation}
where $\widetilde{H}\equiv i \sigma_2 H$ is the conjugate $SU(2)_L$ Higgs doublet, $y_{\nu,N,\chi}$ are Yukawa couplings, $\alpha=1,2,3$ is a flavor index and we omitted the flavor indices in the dark sector for simplicity. There is an explicit Majorana mass terms $\mu_N$ allowed by gauge symmetry that can be introduced at tree level while the gauge symmetry does not allow for a Majorana mass term for $N_R'$.
 Since $\mu_N$ is the only source of explicit lepton-number violation in the Lagrangian, this parameter is technically natural in the sense that since its running vanishes in the limit where this parameter goes to 0, a small chosen value of $\mu_N$ should remain small at all scales~\cite{tHooft:1979rat}. Notice that a term $\mathcal{L} \supset \Phi \overbar{\chi}_R \omega_L$ could be included but one can always redefine the fields to a basis where this term is absent. In the following we will analyse the implications of the individual parts of the Lagrangian separately.

\subsection{The scalar sector}
From Eq.~(\ref{eq:Lscalar}) the vacuum expectation values are given by:
\begin{equation}
    v_\phi=\dfrac{\mu _{\phi }}{\sqrt{\lambda _{\phi }}} \quad \text{and} \quad v_h=\dfrac{\mu _{h }}{\sqrt{\lambda _{h }}}~,
\end{equation}
in the limit where the quartic mixing is negligible. In this limit the scalar mass mixing in the basis $(h,\phi)$ reads
\begin{equation}
M_0^2=\left(\begin{array}{cc}
        \lambda_h v_h^2 & \lambda_{\phi h} v_h v_\phi/2 \\
       \lambda_{\phi h} v_h v_\phi/2 &  \lambda_\phi v_\phi^2
    \end{array}\right)~,
\end{equation}
which implies a mixing angle between the gauge and mass eigenstates with an angle $\alpha$
\begin{equation}
\left(\begin{array}{c} h_1 \\ h_2 \end{array}\right)=\left( \begin{array}{cc} \cos \alpha &  - \sin \alpha  \\   \sin \alpha & \cos \alpha \\  \end{array} \right)  \left(\begin{array}{c} h \\ \phi \end{array}\right)~,
\end{equation}
defined as
\begin{equation}
    \tan (2\alpha) = \dfrac{\lambda_{\phi h }v_h v_\phi}{\lambda_\phi v_\phi^2-\lambda_h v_h^2}~,
\end{equation}
where $h_1 \simeq h$ ans $h_2 \simeq \phi$ are the mass eigenstates. Therefore for simplicity in the following we denote the mass eigenstates as $h$ and $\phi$. In the limit where the mixing angle is small, the mass of the new scalar $\phi$ is given by:
\begin{equation}
    m_\phi=\sqrt{2\lambda_\phi} v_\phi~,
\end{equation}
The kinetic terms of the complex scalar will give rise to the mass term for the $Z^\prime$ as
\begin{equation}
        \mathcal{L}_\text{scalar}\supset|D^\mu \Phi|^2\supset \dfrac{g_{Z^\prime}^2}{2}\Big(v_\phi+\phi\Big)^2 Z^{\prime}_\mu  Z^{\prime \mu}\supset \dfrac{m_{Z^\prime}^2}{2} Z^{\prime}_\mu  Z^{\prime \mu}~,
\end{equation}
with $m_{Z^\prime}=g_{Z^\prime} v_\phi$. In Sec.~\ref{sec:pheno} will be summarize the constraints on the scalar and investigate its role to obtain the correct DM relic density.

\subsection{The gauge sector}
\label{sec:thegaugesector}

The kinetic and mass terms for the $Z$ and $Z'$ gauge bosons can be diagonalized by
\be
\left(\begin{array}{c} B_\mu \\ W^3_\mu \\ Z'_\mu \end{array}\right)=\left( \begin{array}{ccc} c_W & -s_W c_\zeta +t_\xi s_\zeta  & -s_W s_\zeta-t_\xi c_\zeta  \\   s_W & c_W c_\zeta & c_W s_\zeta   \\  0 & -s_\zeta/c_\xi & c_\zeta/ c_\xi   \end{array} \right)  \left(\begin{array}{c} A_\mu \\ Z_{1\mu} \\ Z_{2\mu} \end{array}\right)~,
\ee
where $(B_\mu, W^3_\mu,Z'_\mu)$ are hypercharge, electroweak and dark gauge fields, $(A_\mu, Z_{1\mu},Z_{2\mu})$ are mass eigenstates, and $s_W\equiv \sin\theta_W, c_W\equiv \cos\theta_W$, etc. In the regime of small kinetic mixing $\xi \ll 1$, and large mass hierarchy $m_Z \gg m_{Z^\prime}$, the mass eigenvalues of $Z$-boson and dark photon are $m_1\approx m_Z$, $m_2\approx m_{Z'}$
and the mixing angle $\zeta$ between the $Z$-boson and the dark photon is given by
\be
\tan( 2\zeta) =\frac{m^2_Z s_W \sin 2\xi}{m^2_{Z'}-m^2_Z}~.
\ee 
which reduces to $\zeta \simeq -\xi  s_W $ for $m_{Z^\prime} \ll m_Z$. In this regime the interactions terms between electroweak fields and various current reduce to 
\bea
{\cal L} \supset e A_\mu J^\mu_{\rm EM} + Z_{1\mu} \bigg[ \frac{e}{s_Wc_W} J^\mu_Z +g_{Z'} \varepsilon t_W J^\mu_{Z'} \bigg]+Z_{2\mu}  \bigg[- e \varepsilon J^\mu_{\rm EM}+g_{Z'} J^\mu_{Z'} \bigg]~,
\label{eq:IntCurrentsPureKinMixing}
\eea
where $\varepsilon \simeq c_W \xi$, $J_\text{EM}^\mu$ and $J_Z^\mu$ are the usual SM ElecroMagnetic (EM) and $Z$ boson current and $J_{Z^\prime}^\mu$ is the current associated to the extra gauge symmetry $U(1)^\prime$~\footnote{A term $\dfrac{\varepsilon e}{c_W^2} \dfrac{m_{Z^{\prime}}^2}{m_{Z^{\prime}}^2-m_{Z}^2} J_Z^\mu Z^\prime_{2\mu}$ in Eq.~(\ref{eq:IntCurrentsPureKinMixing}) should also present at leading order in $\varepsilon$ but is negligible in the limit $m_{Z^\prime} \ll m_Z$}.
We recover the usual results, $Z_{2\mu}\simeq Z^\prime_\mu$ does only couple to the EM current and  $Z_{1\mu}\simeq Z_\mu$ couples to the $Z^\prime$ current at leading order in $\varepsilon$~\cite{Fayet:1989mq,Fayet:2016nyc}.\\
Additionally, kinetic mixing between the neutral gauge bosons can be induced at loop level if there are fields  charged under both gauge groups~\cite{Holdom:1985ag, Cheung:2009qd}. This is not the case in our model as no SM field is charged under the dark $U(1)'$. Nevertheless, due to the mixing of scalars, and the mixing of the neutrino a loop-level contribution to the kinetic mixing is generated.
An order of magnitude estimate of the size of this contribution leads to
$\varepsilon_\text{loop}\sim (10^{-2}-10^{-1})\times g_{Z'}\times \vartheta^2$ where $\vartheta$ stands representative for either the scalar mixing angle or the active-sterile neutrino mixing angle~\cite{Gherghetta:2019coi}. In Sec.~\ref{sec:pheno} we derive constraints on the viable values of the parameter $\varepsilon$.

\subsection{The dark sector}
\label{sec:modelthedarksector}

The dark sector of the model contains $N_R,N_R'$, the DM candidate and the massless fermion $\omega$. A study of the phenomenology of the neutrinos in the hidden sector will be postponed to the next section. In our model the DM candidate is the Dirac fermion $\chi$, whose components are $\chi_{L,R}$, formed after $\Phi$ acquires a non-vanishing vev and the new $U(1)'$ symmetry is broken. Its mass is generated by the Yukawa term
\begin{equation}
    \mathcal{L}\supset -y_\chi \Phi \overbar{\chi}_R \chi_L +\text{h.c.}= - \dfrac{y_\chi}{\sqrt{2}} (\phi+v_\phi) \overbar{\chi} \chi \supset - m_\chi \overbar{\chi} \chi ~,
\end{equation}
with $m_\chi=y_\chi v_\phi /\sqrt{2}$. The DM candidate can interact with the SM and dark sector both via $\phi$ and $Z^\prime$ which mix respectively with the SM Higgs and photon. The coupling between the DM and the light $Z^\prime$ follows from the expansion of the covariant derivative of the DM kinetic terms 
\begin{equation}
    \mathcal{L} \supset g_{Z^\prime} \overbar{\chi} \gamma^\mu(V_\chi-A_\chi \gamma_5) \chi Z^\prime_\mu ~~\text{with}~~V_\chi \equiv \dfrac{q_{\chi _L}+q_{\chi _R}}{2} =\dfrac{9}{2}~~\text{and}~~A_\chi \equiv \dfrac{q_{\chi _L}-q_{\chi _R}}{2} =-\dfrac{1}{2}~,
\end{equation}
where $q_{\chi_L}$ and $q_{\chi _R}$ are respectively the charges of the left and right-handed DM component with respect to the $U(1)'$ gauge symmetry.
The last particle of the dark sector is the Majorana field $\omega \equiv \omega_L+\omega^c_L$ which remain massless and only interacts via gauge interactions. As stated previously, a term like $\mathcal{L} \supset \Phi \overbar{\chi}_R \omega_L$ is allowed by gauge invariance and should be included, but it can be rotated away by field redefinitions. Once such rotation is performed, the Lagrangian exhibits two accidental $\mathbb{Z}_2$ symmetries, one is responsible for the DM stability and the other one ensures that $\omega$ remains massless. Therefore in our setup DM stability is not an \textit{ad-hoc} artefact but is a consequence of the consistency of the gauge symmetry of the model. Moreover, its connection to the presence of this massless state, which will be shown to play a crucial cosmological role, provides a way of testing the viability of the entire construction.

\section{Neutrino masses and mixings}
\label{sec:neutrino}
After the breaking of $U(1)^\prime$, most of the dark sector states will mix. The mass matrix involving SM neutrinos $\nu_L$, and the extra states $N_R$ and $N_R^\prime$ can be written as~\footnote{As 3 families of $\nu_L, N_R, N_{R^\prime}$ are actually present, this matrix should be $9 \times 9$, however we assume flavour-diagonal couplings here for clarity.}
\begin{equation}
    \mathcal{L}^\text{tree}_\text{mass}=\dfrac{1}{2}\left(\begin{array}{ccc}
        \overbar{\nu_L} & \overbar{N_R^c} & \overbar{N_R^{\prime c}} \end{array}\right) \left( \begin{array}{ccc}
    0 & m_D & 0 \\ m_D^\dagger & \mu_N & M_R \\ 0 & M_R^\dagger & 0\end{array}  \right) \left(\begin{array}{c}
        \nu_L^c \\ N_R \\ N_R^{\prime} \end{array}\right) +\text{h.c.}
        \label{eq:massmatrix}
\end{equation}
where we used the notation $m_D=y_\nu v_h/ \sqrt{2}$ and $M_R=y_N v_\phi / \sqrt{2}$. This matrix is actually singular and hence the SM neutrinos will remain massless if only tree-level terms are considered. However, as discussed further on, loop corrections will give rise to non-vanishing neutrino masses. 
After diagonalizing the mass matrix of Eq.~(\ref{eq:massmatrix}), we end up with $N_F$ ($N_F$ is the number of right-handed neutrino families, in our model $N_F=3$) copies of a set of two mass eigeinstates whose eigenvalues are given by
\begin{equation}
    m_{5,4}=\dfrac{\mu_N \pm \sqrt{\mu_N^2+4 \left(M_R ^2+m_D^2\right)}}{2}~.
\end{equation}

Rewriting the interaction terms involving $N_R$ and $N_R^\prime$ in terms of the mass eigenstates (the Majorana fermions $N_{4,5}$) leads to
\begin{equation}
        \mathcal{L} \supset -\dfrac{y_N \phi}{\sqrt{2}} \Big( \overbar{N}_4 N_4+\overbar{N}_5 N_5 \Big)+\dfrac{g_{Z^\prime} Z^\prime_\mu}{4} \Big[ \overbar{N}_5 \gamma^\mu \gamma_5 N_5+\overbar{N}_4 \gamma^\mu \gamma_5 N_4 + 2 i \overbar{N}_4 \gamma^\mu N_5  \Big]
\end{equation}
The connection between the hidden sector of the model and the SM part of the Lagrangian is made through the Yukawa coupling
\begin{equation}
    \mathcal{L}_\text{fermions}=-y_\nu^\alpha \overbar{L_L^\alpha} \widetilde{H} P_R \left( \dfrac{N_5 + i N_4}{\sqrt{2}} \right) +~\text{h.c.} 
\end{equation}
As previously stated, since the mass matrix in Eq.~(\ref{eq:massmatrix}) is singular, the neutrinos are expected to be massless at tree level. However since $\mu_N$ is an explicit lepton number violating terms and no symmetry forbid the SM-like neutrino masses to receive quantum corrections, one expects Majorana mass terms $m^\nu_{\alpha \beta}$ to be generated at the loop level~\cite{Dev:2012sg,Fernandez-Martinez:2015hxa,LopezPavon:2012zg} from diagrams involving a $\mu_N$ insertion and a loop of $Z, Z^\prime$ or $h, \phi$ and associated Goldstone bosons. Following~\cite{Ballett:2019cqp}, the masses generated at the one loop level can be written as
\begin{equation}
    m_{ij}=\dfrac{1}{4\pi^2} N_F \sum_{k=4}^{5} \Big[C_{ik} C_{jk} \dfrac{m_k^3}{m_Z^2}F(m_k^2,m_Z^2,m_h^2)+D_{ik} D_{jk} \dfrac{m_k^3}{m_{Z^{\prime}}^2}F(m_k^2,m_{Z^{\prime}}^2,m_\phi^2) \Big] ~.
    \label{eq:numasses}
\end{equation}
where $k=4,5$ and $i,j=1,2,3$ denote respectively the new states and the SM-like states in the mass basis and $N_F$ the multiplicity of the new states, in our case $N_F=3$. The loop function $F$ can be defined as
\begin{equation}
    F(a,b,c)\equiv \dfrac{3 \log (a/b)}{a/b-1}+\dfrac{\log (a/c)}{a/c-1}~,
\end{equation}
and 
\begin{equation}
    C_{ik}\equiv \dfrac{g_{Z^\prime}}{4c_W} \sum_{\alpha=e}^\tau U^*_{\alpha i} U_{\alpha k}~, \quad \quad \text{and} \quad \quad  D_{ik}\equiv\dfrac{g_{Z^\prime}}{2}  U^*_{R' i} U_{R' k}~,
\end{equation}
where $\alpha$ denotes the SM-neutrino flavour-eigenstates. These coefficients satisfy the following relations 
\begin{equation}
    \sum_k m_k C_{ik} C_{jk}=0~, \quad \quad \text{and} \quad \quad \sum_k m_k D_{ik} D_{jk} =0~.
\end{equation}
In the limit where the SM contribution can be neglected
\begin{equation}
    m_{ij}=\dfrac{g^{\prime 2}}{32\pi^2} N_F U^*_{Ri} U^*_{Rj}\dfrac{m_5}{m_{Z^{\prime}}^2} \Big[ m_5^2F(m_5^2,m_{Z^{\prime}}^2,m_\phi^2) -m_4^2 F(m_4^2,m_{Z^{\prime}}^2,m_\phi^2)\Big]~.
\end{equation}
In the limit where $M_R \gg \mu_N$, the mixing angles between flavour and mass eigenstates are
\begin{equation}
    U_{\alpha 4,5}\simeq \dfrac{m_D}{\sqrt{2} M_R}~, \quad U_{R i}\simeq U_{R^\prime i}\simeq \dfrac{m_D}{M_R}~, \quad U_{R 4,5}\simeq U_{R^\prime 4,5}\simeq \dfrac{1}{\sqrt{2}}~,
\end{equation}
where $\alpha=e,\mu,\tau$.
In the limit where $M_R \gg m_{Z^\prime},m_{\phi}$, the one-loop generated mass for the heaviest SM-like neutrino state can be approximated as~\footnote{Assuming $\nu_3$ to be the heaviest SM-like state.}
\begin{equation}
    m_{3}=N_F\frac{g_{Z^\prime}^{2} m_D^2 \mu _N }{16 \pi ^2 M_R ^2} \left(3+\frac{m_\phi^2}{m_{Z^\prime}^2}\right)~,
\end{equation}
while in the limit where $M_R \ll  m_{Z^\prime},m_{\phi}$ it is given by
\begin{equation}
    m_{3}=N_F\frac{g_{Z^\prime}^{2} m_D^2 \mu_N }{16 \pi ^2 m_{Z'}^2} \Big[ 3 \log \left( \dfrac{m_{Z^\prime}^2}{M_R^2} \right) + \log \left( \dfrac{m_{\phi}^2}{M_R^2} \right)-4 \Big]~.
\end{equation}
The SM contribution to the neutrino masses is given by
\begin{equation}
    m_{ij}=N_F\dfrac{\alpha_W}{16\pi^2} \sum_{\alpha,\beta=e,\mu,\tau} U^*_{\alpha i} U^*_{\beta j} U_{\alpha 5} U_{\beta 5} \dfrac{m_5}{m_{W}^2} \Big[ m_5^2F(m_5^2,m_{Z}^2,m_h^2) -m_4^2 F(m_4^2,m_{Z}^2,m_h^2)\Big]~,
\end{equation}
and  assuming $m_D, \mu_N \ll M_R$ we have the following approximated formula
\begin{equation} 
m_3 \simeq N_F \dfrac{\alpha_{W} \mu_N  m_D^2 }{16 \pi m_W^2} (U^*_{e3}+U^*_{\mu 3}+U^*_{\tau 3})^2 \left[\log \left(\dfrac{m_h^2}{M_R ^2}\right)+3 \log \left(\dfrac{m_Z^2}{M_R ^2}\right)-4\right]~.
\end{equation}
In order to obtain neutrino masses below 0.1 eV in the case of $M_R\gg m_{Z'}, m_\phi$, the Dirac mass term $m_D$ needs to be between $10^{-3}-10$ GeV
taking $m_{Z'}\sim m_\phi\sim 100\times M_R$ between $10^{-2}-10^2$ GeV, and setting $g_{Z'}\sim 0.1$ and $\mu_N\sim 0.1$ GeV.
In the opposite limit if $M_R\ll m_{Z'}, m_\phi$ the Dirac mass term $m_D$ also needs to be between $10^{-3}-10$ GeV
taking $M_R\sim m_{Z'}\times 100 \sim m_\phi\times 100$ between $10^{-2}-10^2$ GeV, setting $g_{Z'}\sim 0.1$ and $\mu_N\sim 0.1$.
For the SM contribution to the neutrino mass, $m_D$ needs to be between $10^{-2}-10$ GeV for $\mu_N$ between $10^{-2}-10^2$ GeV and $M_R\sim 10$ GeV.\\
The phenomenology of the heavy neutrinos has been studied in detail in the model from~\cite{Ballett:2019cqp} which exhibits the same neutrino sector. Despite the fact that in our model the dark sector is enlarged, the same constraints on the heavy neutrino masses and mixings apply as the presence of $\chi$ and $\omega$  do not affect the neutrino phenomenology.
Here we summarise the main constraints and detection prospects of the heavy neutrinos.
The mixing of the sterile neutrinos with muon neutrinos  provides  the most sensitive avenue to test the additional neutrinos.Constraints from meson decay peak searches~\cite{Yamazaki:1984sj,Artamonov:2014urb,CERNNA48/2:2016tdo}, beam dump  experiments~\cite{Bernardi:1985ny,Bergsma:1983rt,Badier:1986xz,Vaitaitis:1999wq, CooperSarkar:1985nh, Astier:2001ck} and collider experiments~\cite{Abreu:1996pa,Akrawy:1990zq,Sirunyan:2018mtv} constrain  $|U_{\mu4}|^2$, $|U_{\mu5}|^2\lesssim 10^{-5}$ for  $m_4,~m_5$ between $10^{-2}-10^1$ GeV, for $m_4,~m_5$ around $0.2-0.5$ GeV peak searches  constrain the mixing angles down to $ |U_{\mu4}|^2$, $|U_{\mu5}|^2\lesssim 10^{-8}$.
Current and future neutrino experiments can probe a  large region of the parameter space as it has been shown in~\cite{Ballett:2019cqp}. In particular  the Short-Baseline Neutrino program (SBN)~\cite{Ballett:2016opr} and the Deep Underground Neutrino Experiment (DUNE) near detector~\cite{Ballett:2018fah,Ballett:2019bgd} with heavy neutrinos in decay-in-flight searches can probe the parameter space.  Also the NA62 Kaon factory operating in beam dump mode~\cite{Drewes:2018irr}, and the dedicated beam dump experiment Search for Hidden Particles (SHiP)~\cite{Bonivento:2013jag,Alekhin:2015byh} will cover a  large region of parameter space from $400$ MeV to $\lesssim6$ GeV.
The bounds on the mixing with electron neutrinos are of similar order as in the muon sector while the mixing with tau neutrinos is poorly constraint and a large region of parameter space is allowed~\cite{Atre:2009rg,Deppisch:2015qwa, Bolton:2019pcu}.

Further detection prospects of the heavy neutrinos are via their invisible decays which have not been taken into account for the above mentioned constraints.
In our model the dominant decays of the heavy neutrinos are similar as in the type I seesaw, namely via the active-sterile mixing angle  suppressed  charged and neutral current interactions mediated by the SM gauge bosons.

Finally, as will be discussed in Sec.~\ref{sec:pheno},
cosmological bounds on the sterile neutrinos can be evaded if their cosmological densities are negligible during Big Bang Nucleosynthesis (BBN) and hence do not leave any imprint in the number of relativistic species in the early Universe.

\section{Thermal history of the dark sector}
\label{sec:ds}
As described in Sec.~\ref{sec:model} our model predicts a massive stable fermion which is our DM candidate as well as a massless fermion. Both particles play an essential cosmological role in this model, that is investigated into details in this section. At the beginning, we discuss about the various possible annihilation modes involved in the DM density production. In the following subsections we investigate the parameter space allowing the hidden-sector to thermalize with the SM plasma. We estimate the freeze-out temperature, for which both sectors finally decouple, and its impact on the effective number of relativistic degrees-of-freedom, carried out by the massless state, confronting its values to current and future bounds and emphasizing its connection to the recently established Hubble tension.
\subsection{Dark Matter relic density}
The most recent Planck analysis determined a precise estimate of the dark-matter relic density at the present day $\Omega_\chi h^2 =0.11933 \pm 0.00091$ based on TT, TE, EE + lowE + lensing + BAO~\cite{Aghanim:2018eyx}. In the context of WIMPs, assuming the DM in thermal equilibrium with the SM particles in the early universe~\footnote{This condition will be checked in the following subsections.}, the relic density is related to the velocity-averaged annihilation cross section $\Omega_\chi h^2 \propto 1/\langle \sigma v \rangle$ and the correct dark matter abundance can be achieved for $\langle \sigma v \rangle \sim 3 \times 10^{-26}~\text{cm}^3~\text{s}^{-1}$~\cite{Steigman:2012nb,Arcadi:2017kky}. The status of WIMP DM have been studied in recent works~\cite{Escudero:2016gzx,Arcadi:2017kky,Blanco:2019hah,Roszkowski:2017nbc,Balazs:2017ple} that highlighted the complementarity from collider, direct and indirect searches, pushing the remaining allowed parameter space of models where DM annihilates into or via SM states, towards restrained corners or higher mass scales. Therefore in our case we focus in a WIMP regime where DM achieve thermalization with SM particles but annihilate mostly into particles of the dark sector and subdominantly to SM particles through various mixings~\footnote{Recent works~\cite{Chu:2011be,Hambye:2019dwd} investigated mixed solutions where dark-sector particles can be produced by freeze-in and thermalize among themselves without ever equilibrating with the SM particle content. Such considerations could be applied in our setup but are out of the scope of this paper.}. In the following section we present the possible annihilation channels and corresponding cross sections as well as some values of the parameters allowing to account for the observed dark matter abundance. To estimate the cross section we perform the usual expansion in terms of powers of the mean DM velocity $\bar{v}_\chi$ evaluated at the DM freeze-out temperature $x_\text{F}=m_\chi/T_\text{F}\simeq 20$~\cite{Gondolo:1990dk,Jungman:1995df}, which remains valid away from poles or kinematic thresholds~\cite{Griest:1990kh}. In the following we give some analytical expressions for the annihilation channels $\bar{\chi} \chi \to \omega \omega, Z^\prime Z^\prime, \phi \phi, N_{4,5} N_{4,5}, \bar{\psi} \psi$ (where $\psi$ is a SM fermion), at leading order in the mixing angles between $Z,Z^\prime$ and $h,\phi$ :
\begin{figure}[t!]
    \centering
    \includegraphics[width=0.48\linewidth]{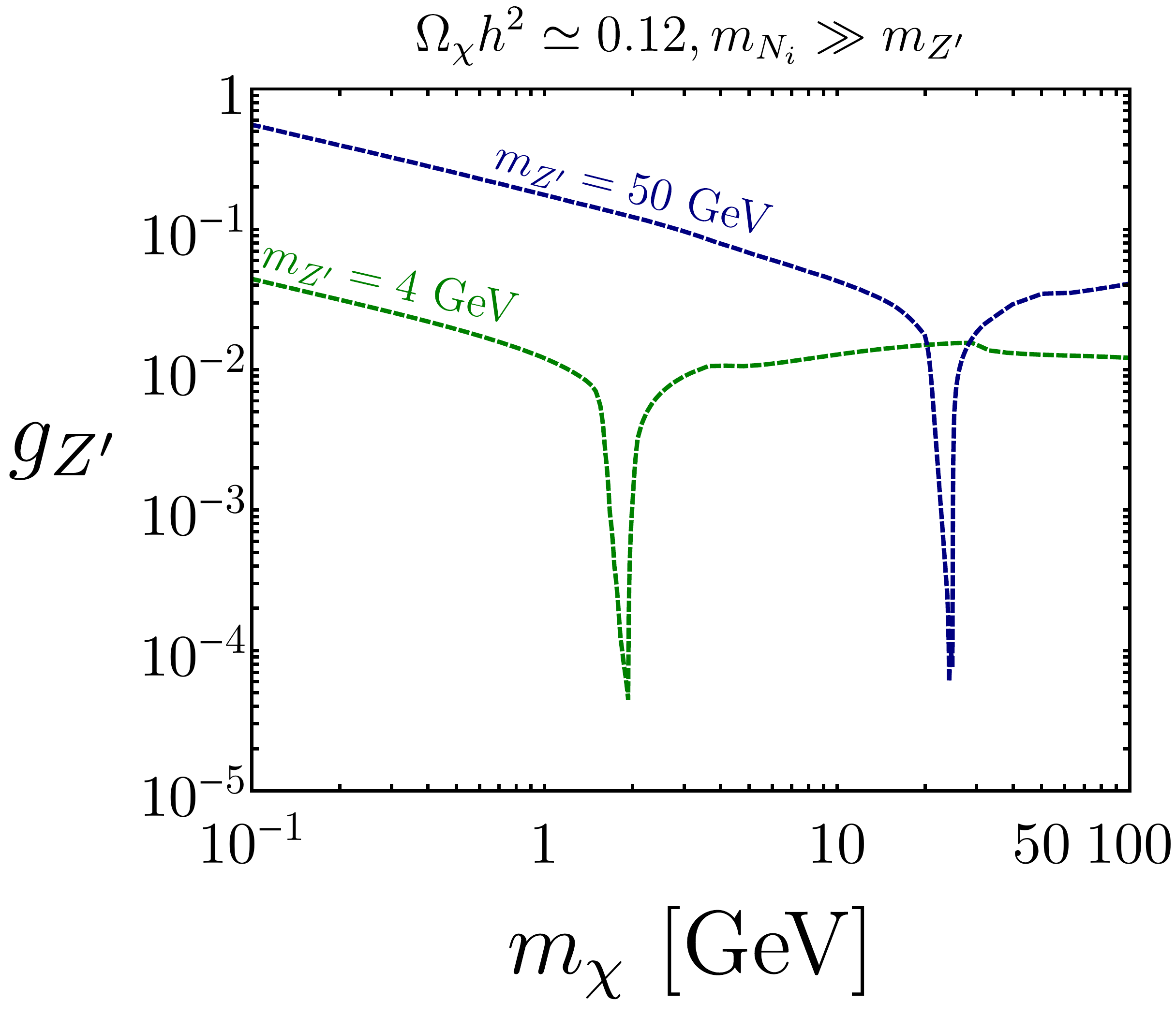}
        \includegraphics[width=0.48\linewidth]{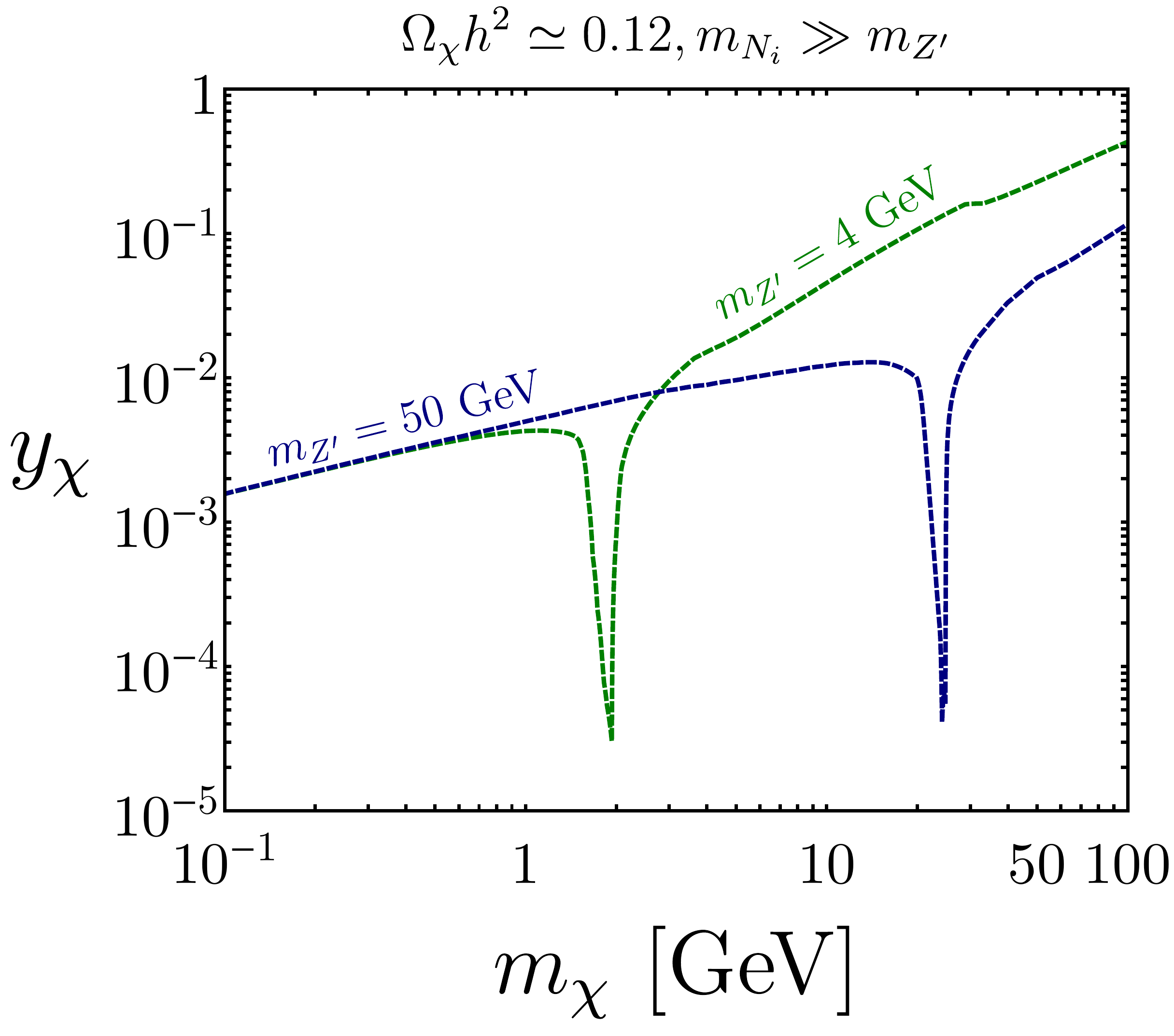}
    \caption{Gauge coupling of the dark sector (left) and DM Yukawa coupling (right)  as a function of the DM mass satisfying the relic density condition, computed numerically using micrOMEGAs~\cite{Belanger:2018mqt}. Masses for the heavy neutrinos are taken to be $m_{N_i}=30$ GeV for $m_{Z^\prime}=4$ GeV and $m_{N_i}=300$ GeV for $m_{Z^\prime}=50$ GeV. Values of the various couplings are computed for each set of parameters $g_{Z^\prime}$ and $m_\chi$, assuming $m_\phi = v_\phi$ and mixing angles to be negligible.}
    \label{fig:gauge_coupling_relic_density}
\end{figure}

\begin{equation}
    \langle \sigma v \rangle_{\bar{\chi}\chi  \rightarrow \omega \omega} \simeq \frac{162 g_{Z'}^4 m_{\chi }^2 }{\pi (m_{Z'}^2-4 m_{\chi }^2)^2} \simeq 3 \times 10^{-26}~\text{cm}^3\text{s}^{-1}~\left( \dfrac{g_{Z^\prime}}{4\times 10^{-2}} \right)^4 ~\left( \dfrac{m_\chi}{10~\text{GeV}} \right)^2 \left( \dfrac{50~\text{GeV}}{m_{Z^\prime}} \right)^4 ~,
\label{eq:chichitoomegaomega}
\end{equation}
where we took the limit $m_\chi \ll m_{Z^\prime}$ in the second part of the previous expression,
\begin{equation}
            \langle \sigma v \rangle_{\bar{\chi}\chi  \rightarrow Z^\prime Z^\prime}  \simeq \frac{81 g_{Z'}^4}{32 \pi  m_{Z'}^2}  \simeq  3 \times 10^{-26}~\text{cm}^3\text{s}^{-1}~\left( \dfrac{g_{Z^\prime}}{4\times 10^{-2}} \right)^4 \left( \dfrac{50~\text{GeV}}{m_{Z^\prime}} \right)^2~,
\end{equation}
\begin{equation}
     \langle \sigma v \rangle_{\bar{\chi}\chi  \rightarrow \phi \phi} \simeq  \bar{v}_{\chi }^2  \frac{3 m_{\chi }^2  }{128 \pi  v_{\phi }^4} \simeq 3 \times 10^{-26}~\text{cm}^3\text{s}^{-1} ~\left( \dfrac{m_\chi}{100~\text{GeV}} \right)^2~\left( \dfrac{200~\text{GeV}}{v_\phi} \right)^4~,
\end{equation}
\begin{equation}
     \langle \sigma v \rangle_{\bar{\chi}\chi  \rightarrow N_i N_i} \simeq   
    N_F \dfrac{81 m_{\chi }^2 g_{Z'}^4}{32 \pi  (m_{Z'}^2-4 m_{\chi }^2)^2}+  N_F \bar{v}_{\chi }^2 \dfrac{m_{i}^2 m_{\chi }^4 }{16 \pi  v_{\phi }^4 (m_{\phi }^2-4 m_{\chi }^2)^2}~,
    \label{eq:chichitoN5N5}
\end{equation}
with $i=4,5$, which correspond respectively to $s-$wave and $p-$wave processes. 
$N_F=3$ is the multiplicity of the heavy neutrino states $N_{4,5}$.
The annihilation into SM fermions is given by
\begin{equation}
     \langle \sigma v \rangle_{\bar{\chi}\chi  \rightarrow \bar{\psi} \psi} \simeq c_\psi Q_\psi^2  \frac{81 e^2 \varepsilon ^2  g_{Z'}^2 m_{\chi }^2}{4 \pi  (m_{Z'}^2-4 m_{\chi }^2)^2}+ c_\psi \bar{v}_{\chi }^2   \frac{\sin^2(\alpha) m_\psi^2 m_{\chi }^4 (m_h^2-m_{\phi }^2)^2}{32 \pi  v_h^2 v_{\phi }^2 (m_h^2-4 m_{\chi }^2)^2 (m_{\phi }^2-4 m_{\chi }^2)^2}~,
\end{equation}
where we neglected the $Z$ propagator, and $c_\psi$ and $Q_\psi$ are respectively the color factor and electric charge of the SM fermion $\psi$. \\
 Essentially, as the decay channel $\bar \chi \chi \to \omega \omega$ is always open and mediated by gauge interactions, as long as the gauge coupling is $g_{Z^\prime} \gtrsim y_\chi$, this process is always dominant or comparable to annihilation channels such as $\bar \chi \chi \to  N_{4,5} N_{4,5}, Z^\prime Z^\prime$ when kinematically allowed. In the limit where the DM Yukawa coupling is dominant over the gauge coupling $y_\chi \gtrsim g_{Z^\prime} $ annihilations to $\bar \chi \chi \to \phi \phi $ and $\bar \chi \chi \to N_{4,5} N_{4,5} $ are the most efficient processes however both processes are velocity and helicity suppressed
 therefore less efficient to achieve the correct relic density, restricting the vev $v_\phi \lesssim 400~\text{GeV}$ to satisfy the correct relic density with a Yukawa coupling not larger than 1. While in the inverse regime $g_{Z^\prime}  \gtrsim y_\chi $, imposing the gauge coupling to be smaller than 1 implies a much larger upper bound on the vev $v_\phi \lesssim 40~\text{TeV}$ not to overclose the universe\footnote{This argument is not valid on resonances as discussed further on.}. However we restrict our analysis to masses of the order of the typical electroweak scale in this work. In order to make more accurate estimations, the relic density is computed numerically using micrOMEGAs~\cite{Belanger:2018mqt} after implementing the model in Feynrules~\cite{Alloul:2013bka}. Numerical results are depicted in Fig.~\ref{fig:gauge_coupling_relic_density} which are in good agreement with analytical approximations in their own validity regimes.  Typically, a gauge coupling of $g_{Z^\prime}\sim 10^{-2}-10^{-1}$ and DM masses of the order of the $1-100$ GeV range can account for the dark matter relic abundance. Notice that on the pole regions, where $m_\chi \sim m_{Z^\prime}/2$, the expected value of the vev $v_\phi=m_{Z^\prime}/g_{Z^\prime}$ can reach values beyond $1-100$ TeV, therefore requiring the DM Yukawa couplings to be relatively small and making the theoretical framework less appealing for such tuned parameters.

\subsection{Thermalization of the dark sector via kinetic mixing}
For the WIMP mechanism described in the previous subsection to be valid, thermalization between the hidden sector and SM must be achieved in the early universe. In this subsection we investigate the possibility of achieving thermalization of both sectors via kinetic mixing $\varepsilon$ between the dark $Z^\prime$ and the SM hypercharge field. At leading order in $\varepsilon$ a simple way of thermalizing the dark sector with the SM is to produce a large population of $Z^\prime$ that will subsequently generate a population of $\omega$ by inverse-decays and annihilations, and then eventually the rest of the dark sector which forms a thermal bath on its own, since $g_{Z^\prime}$ is $\mathcal{O}(10^{-1}-10^{-2})$ and the dark sector fields interact rather strongly with each other. The $Z^\prime$ can be produced by inverse-decay from $\bar{\psi}+  \psi \rightarrow Z^\prime$ or by $2\rightarrow 2$ annihilations such as $\psi + \gamma \rightarrow \psi+ Z^\prime$ where $\psi$ denote any charged particle of the SM. As the temperature of the universe decreases, interactions between both sectors become too suppressed and the dark sector freezes-out, leading to an extra contribution to the effective number of relativistic species which is investigated later on in this subsection.

\subsubsection{Production of a $Z^\prime$ population and dark sector from inverse-decay}

\begin{figure}[t]
    \centering
    \includegraphics[width=0.48\linewidth]{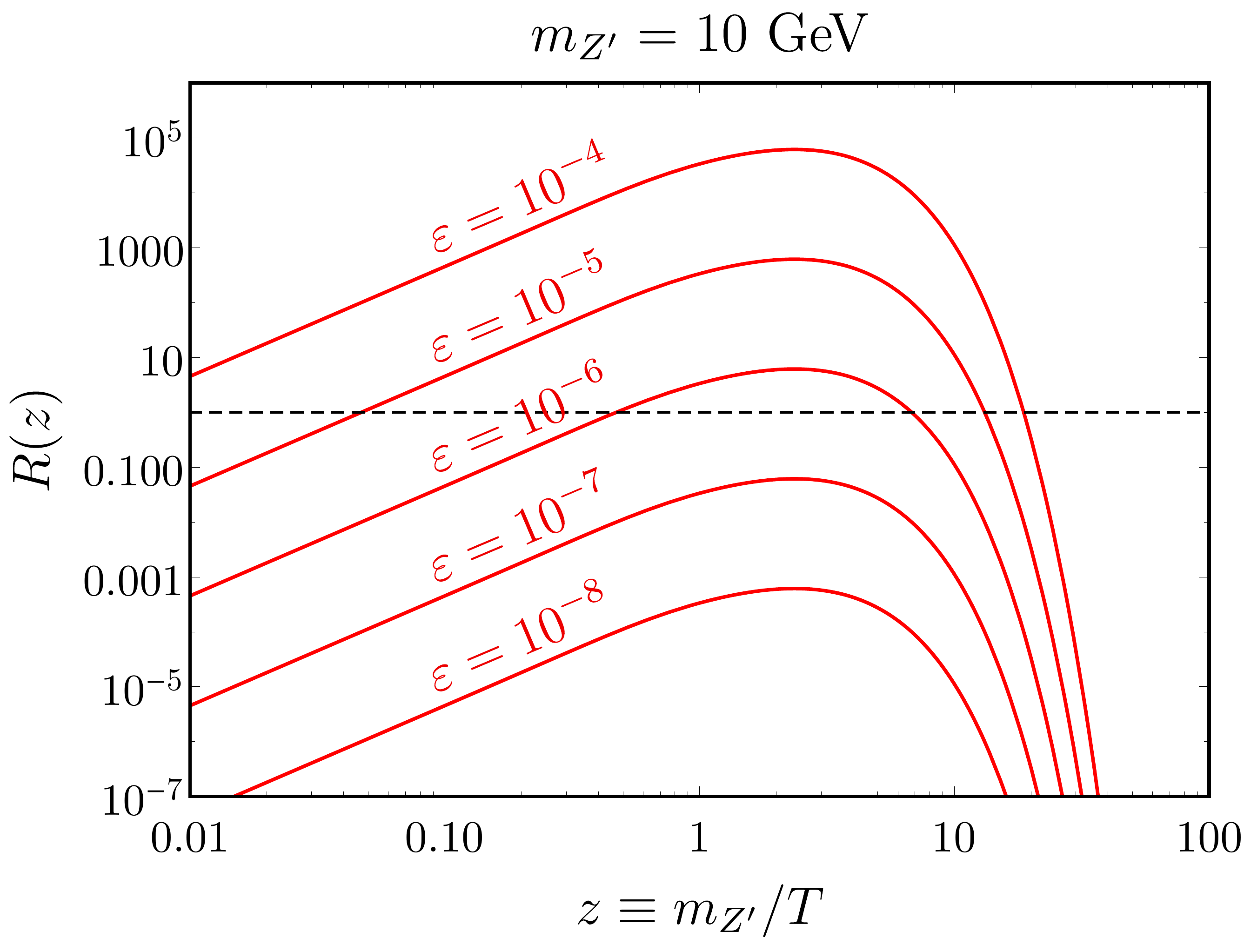}
        \includegraphics[width=0.48\linewidth]{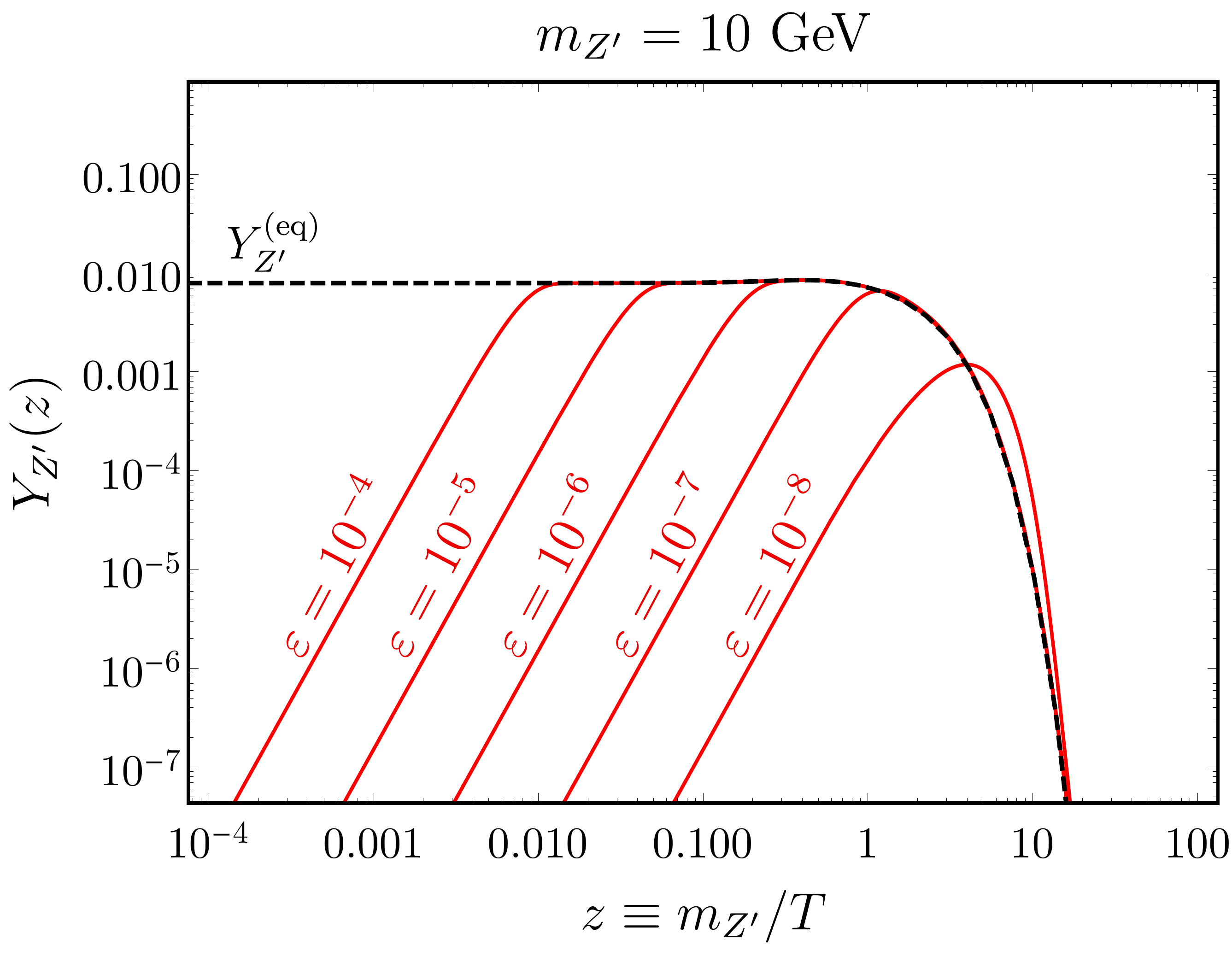}
    \caption{Dimensionless production rate induced by inverse-decay defined in Eq.~(\ref{eq:rateinvdecayZprime}) on the left and numerical solution of the Boltzmann equation of Eq.~(\ref{eq:full-boltz}) on the right for a given set of parameters.}
    \label{fig:production_and_yield_Zp}
\end{figure}

We start our discussion by considering the full Boltzmann equation for the $Z^\prime$ number density $n_{Z^\prime}$, assuming only  the process $\bar\psi(p_1) + \psi(p_2) \leftrightarrow Z^\prime (p_3)$ to be active, which is given by:  
\begin{equation}
    \begin{split}
\frac{\diff n_{Z^\prime}}{\diff t} +3H n_{Z^\prime} =  \int \prod_{i=1}^3 \frac{\diff^3 \vec p_i}{(2\pi)^3 2 E_i} & \left[    |{\cal A}_{\psi\bar \psi \to Z^\prime}|^2 f_1(\vec p_1) f_2(\vec p_2) \right. \\ & \left.  - |{\cal A}_{Z^\prime \to \psi\bar \psi}|^2 f_3(\vec p_3)   \right] (2\pi)^4 \delta^4(p_1+p_2 - p_3) ~,       
    \end{split}
    \label{eq:Boltzmann_inversdecay}
\end{equation}
where $|{\cal A}_{Z^\prime \to \psi\bar \psi}|^2$ is the squared matrix element summed over spin states and $f_i(\vec p_i)$ is the phase space distribution of the particle $i$ involved in the process. As detailed in Sec.~\ref{sec:appendix_collision_terms}, by considering all possible charged SM fermions $\psi$, this equation can be written in terms of the variable $z \equiv m_{Z^\prime}/T$ and dimensionless yield $Y_{Z^\prime} \equiv n_{Z^\prime}/s$ where $s$ is the entropy density as
\be 
\label{eq:full-boltz}
 \frac{\diff Y_{Z^\prime}}{\diff z}  = \sum_\psi  \frac{   \Gamma_{Z^\prime \to \psi\bar \psi}  }{ H(z)  z  } \frac{ K_1(z) }{  K_2(z) }   \left[  Y_{Z^\prime}^{\rm (eq)}(z)  -    Y_{Z^\prime}(z) \right]~,
 \ee
where $Y_{Z^\prime}^{\rm (eq)}(z)$ is given in Sec.~\ref{sec:appendix_collision_terms}, $H$ is the Hubble expansion rate and $K_{1,2}$ are modified Bessel functions of the second kind. The partial width $\Gamma_{Z^\prime \to \psi\bar \psi} $ is given by:
\begin{equation}
   \Gamma_{Z^\prime \rightarrow \bar{\psi}\psi}=   c_{\psi } Q_{\psi }^2 \frac{e^2 \varepsilon ^2 (2 m_{\psi }^2+m_{Z'}^2)}{12 \pi  m_{Z'}}  \sqrt{1-\frac{4 m_{\psi }^2}{m_{Z'}^2}} ~,
   \label{eq:decayZptoSM}
\end{equation}
with $c_\psi$ being a color factor and $Q_\psi$ the electric charge of the SM fermion $\psi$.
Note that this equation accounts for both production and decay processes. We define the dimensionless production rate in the freeze-in regime\footnote{Defined as the regime in which the backreaction term in Eq.~(\ref{eq:Boltzmann_inversdecay}) can be neglected.} as
\begin{equation}
    R(z)\equiv  \sum_\psi  \frac{   \Gamma_{Z^\prime \to \psi\bar \psi}  }{ H(z)  z  } \frac{ K_1(z) }{  K_2(z) } Y_{Z^\prime}^{\rm (eq)}(z) ~,
    \label{eq:rateinvdecayZprime}
\end{equation}
whose numerical evaluation is represented in the left panel of Fig.~\ref{fig:production_and_yield_Zp} which shows that the maximum production rate occurs at $z \sim \mathcal{O}(1)$ and this statement is independent on the specific parameters. In the right panel of Fig.~\ref{fig:production_and_yield_Zp}, we depicted numerical solutions of the Boltzmann equation given some values of the kinetic mixing and $m_{Z^\prime}$. From both panels of Fig.~\ref{fig:production_and_yield_Zp}, one can see that a $Z^\prime$ will reach a thermal equilibrium state with the SM particles at $z=1$ when the rate $R(z)$ is roughly larger than 1 at $z=1$ corresponding to the following condition:
\begin{equation}
   \sum_\psi  \Gamma_{Z^\prime \to \psi\bar \psi}    \gtrsim \left. \dfrac{H(z)z}{Y_{Z^\prime}^{\rm (eq)}(z) } \frac{ K_2(z) }{  K_1(z) } \right|_{z=1} ~.
  \label{eq:condition_Zprime_thermalization}
\end{equation}
For $m_{Z^\prime}=10~\text{GeV}$ this condition corresponds to $\varepsilon \gtrsim 6 \times 10^{-7}$ which might be slightly conservative compared to the value determined graphically from the right panel of Fig.~\ref{fig:production_and_yield_Zp} $\varepsilon \gtrsim  10^{-7}$, but more realistic than the naive estimate obtained by comparing the Hubble rate to the $Z^\prime$ decay width $\Gamma_{Z^\prime}  \gtrsim  H(z=1)$, which gives $\varepsilon \gtrsim 3 \times 10^{-8}$, more than one order of magnitude smaller then the estimate made by using Eq.~(\ref{eq:condition_Zprime_thermalization}). \\
However, since the gauge coupling $g_{Z^\prime}$ is relatively large $g_{Z^\prime}\gg \varepsilon$, the $Z^{\prime}$  population should produce a large number of dark fermions before even reaching thermal equilibrium with the SM particles. The corresponding interaction terms should be taken into account in   Eq.~(\ref{eq:Boltzmann_inversdecay}) but would require a dedicated analysis, beyond the scope of this paper. For simplicity we consider that the conservative condition stated in Eq.~(\ref{eq:condition_Zprime_thermalization}) should be sufficient to account for the effects of dark sector interactions and ensure that thermalization between the entire dark sector and the SM is achieved.

\subsubsection{Production of the dark sector by $2 \rightarrow 2$ annihilations}
\label{sec:producing_dark_sector_via_2to2}
Another possibility of producing the dark sector is to consider $2\rightarrow 2$ annihilations such as $e^\pm \gamma\rightarrow e^\pm Z^\prime$ or $e^+ e^- \rightarrow \omega \omega$ which are the dominant $2 \to 2$ processes at leading order in $\varepsilon$.\footnote{Similar processes like $e^{\pm} \omega\to e^{\pm} \omega$ are of the same order but for the sake of simplicity we consider only $e^\pm \gamma\rightarrow e^\pm Z^\prime$ or $e^+ e^- \rightarrow \omega \omega$ for the estimates below.}
section With a similar idea than in the previous subsection, since $g_{Z^\prime}\gg \varepsilon$, if one particle of the dark sector thermalizes with the SM, we expect the rest of the relativistic species of the dark sector to thermalize as well. In this section we compare the efficiencies of these $2 \to 2$ processes with the inverse-decay production. The Boltzmann equation associated to the process $e^\pm(p_1)+\gamma(p_2)\rightarrow e^\pm(p_3)+Z^\prime(p_4)$ is given by 
\begin{equation}
    \dfrac{\diff n_{Z^\prime}}{\diff t}+3 H n_{Z^\prime}=\left. \dfrac{\delta n_{Z^\prime}}{\delta t}  \right|_{e^\pm \gamma \to e^\pm Z^\prime}-\left. \dfrac{\delta n_{Z^\prime}}{\delta t}  \right|_{e^\pm Z^\prime \to e^\pm \gamma}~,
\end{equation}  
where terms on the right-hand side are collision rates corresponding to the reaction $e^\pm\gamma\rightarrow e^\pm Z^\prime$ and the reverse process.  
Assuming the initial abundance of $Z^\prime$ to be negligible the reverse process is essentially absent in the early universe. The production term is given by
\begin{equation}
\left. \dfrac{\delta n_{Z^\prime}}{\delta t}  \right|_{e^\pm \gamma \to e^\pm Z^\prime}= 2 \int \prod_{i=1}^4 \frac{\diff^3 \vec p_i}{(2\pi)^3 2 E_i} (2\pi)^4  f_1(\vec p_1) f_2(\vec p_2) |{\cal A}_{e^\pm \gamma \to e^\pm Z^\prime}|^2 \delta^4(p_1+p_2- p_3 -p_4 )~,
\label{eq:collision_term_egamma_to_eZprime}
\end{equation}
whose expression is given in Appendix~\ref{sec:appendix_collision_terms}. The Boltzmann equation can be expressed in terms of the dimensionless rate
\begin{equation}
R_{e^\pm \gamma \to e^\pm Z^\prime}(z)\equiv\dfrac{1}{H(z)s(z)z} \left. \dfrac{\delta n_{Z^\prime}}{\delta t}  \right|_{e^\pm \gamma \to e^\pm Z^\prime}~,
\end{equation}
and the $Z^\prime$ yield $Y_{Z^\prime}$ as 
\begin{equation}
    \dfrac{\diff Y_{Z^\prime}}{\diff z}=R_{e^\pm \gamma \to e^\pm Z^\prime}(z)\simeq 1.3 \times \left( \dfrac{\varepsilon}{10^{-5}} \right)^2 \left( \dfrac{10~\text{GeV}}{m_{Z^\prime}} \right)~,
    \label{eq:rate2to2Zprime}
\end{equation}
in the relativistic limit, with $z \equiv m_{Z^\prime}/T$. In a similar way, we can write a Boltzmann equation for the $\omega$ yield $Y_\omega$ production in the process $e^+ e^- \rightarrow \omega \omega$ as:
\begin{equation}
    \dfrac{\diff Y_{\omega}}{\diff z}=R_{e^+ e^- \to \omega \omega}(z)\simeq 1.5 \times 10^{-3} \left( \dfrac{\varepsilon}{10^{-5}} \right)^2 \left( \dfrac{g_{Z^\prime}}{10^{-2}} \right)^2  \left( \dfrac{10~\text{GeV}}{m_{Z^\prime}} \right)~.
        \label{eq:rate2to2omega}
\end{equation}
A comparison of the rates in Eq.~(\ref{eq:rate2to2Zprime}) and in Eq.~(\ref{eq:rate2to2omega}) with the rate of inverse-decay processes of Eq.~(\ref{eq:rateinvdecayZprime}) 
shows that inverse-decays dominate by roughly two orders of magnitude in the infrared regime $z\simeq 1$ and therefore are expected to be the dominant production processes. 
Notice that both rates from $2 \to 2$ processes given in Eq.~(\ref{eq:rate2to2Zprime}) and in Eq.~(\ref{eq:rate2to2omega}) are enhanced for low $m_{Z^\prime}$ mass while inverse-decays are suppressed. 
Consequently, the $2 \to 2$ processes and the inverse-decays would typically have the same efficiency for $m_{Z^\prime} \lesssim \text{GeV}$ therefore in the following we only consider $Z^\prime$ inverse-decay as production mechanism of the dark sector in the kinetic mixing portal case.

\subsubsection{The dark sector freeze-out}

For the typical gauge coupling $g_{Z^\prime}\sim 10^{-2}$ required to achieve the correct relic density, particles of the dark sector form a thermal bath in the early universe and moreover should thermalize with the SM plasma, as described in the previous subsection. However, below a certain temperature $T_\text{FO}$, the interaction rate between the particles of dark thermal bath and the SM plasma becomes smaller than the expansion rate of the universe and the dark sector freezes-out. As $\omega$ particles are massless, one expects them to be the last particles present in the dark thermal bath, playing the same role as photons with respect to the SM plasma. One of the most efficient energy-transfer processes between both sectors are scatterings or annihilations such as $\bar \psi(p_1)+ \psi(p_2)\leftrightarrow \omega(p_3)+ \omega(p_4)$ at lowest order in $\varepsilon$. The freeze-out temperature of the dark sector can be estimated by considering the Boltzmann equation relating the time evolution of the $\omega$ energy density $\rho_\omega$ to the energy-transfer rate of this process:
\begin{equation}
  \dfrac{\diff \rho_{\omega}}{\diff t}+4 H \rho_{\omega}\supset \sum_\psi   \left. \dfrac{\delta \rho_\omega}{\delta t}  \right|_{\bar \psi \psi \to \omega \omega}~,
    \label{eq:Boltzmann_energydensity}
  \end{equation}
with 
\begin{equation}
      \left. \dfrac{\delta \rho_\omega}{\delta t}  \right|_{\bar \psi \psi \to \omega \omega}   \equiv \int \prod_{i=1}^4 \frac{\diff^3 \vec p_i}{(2\pi)^3 2 E_i}  E_1 f_1(\vec p_1) f_2(\vec p_2)  |{\cal A}_{\bar{\psi}\psi\rightarrow \omega \omega}|^2 (2\pi)^4 \delta^4(p_1+p_2- p_3 -p_4 )~.
          \label{eq:collision_term_psipsitoomegaomega}
\end{equation}
This term represents the energy-transfer rate for the reaction $\bar \psi \psi \rightarrow \omega \omega$ whose expression is given in Appendix~\ref{sec:appendix_collision_terms}. The reverse process should also be considered in this equation and its energy transfer is precisely opposite when $\omega$ and $\psi$ are in thermal equilibrium. The freeze-out temperature $T_\text{FO}$ of the dark sector can be determined by considering the moment where the right-hand side of Eq.~(\ref{eq:Boltzmann_energydensity}) becomes smaller than the Hubble expansion term on the left-hand side. Using this condition gives the following expression for the freeze-out temperature:
\begin{equation}
    T_\text{FO}\simeq 1~\text{GeV} \left( \dfrac{m_{Z^\prime}}{10~\text{GeV}}\right)^{4/3} \left( \dfrac{10^{-5}}{\varepsilon}\right)^{2/3} \left( \dfrac{5\times 10^{-2}}{g_{Z^\prime}}\right)^{2/3}~,
    \label{eq:TFO}
\end{equation}
in the limit where the typical momentum transfer in the reaction is much smaller than $m_{Z^\prime}$~\footnote{If this condition is not satisfied, typically for low values of $\varepsilon$, we expect the freeze-out to be close to $T_\text{FO}\sim m_{Z^\prime}$ as resonance effects and $Z^\prime$  (inverse-)decays are frequent enough for such temperatures.}.

\subsubsection{Effective number of relativistic species}
\label{sec:DeltaNeff}
As the universe cools down and drops below the freeze-out temperature, the dark sector decouple from the Standard Model particles. However, the dark-sector energy density still substantially contributes to the Hubble expansion rate. Essentially, as the temperature of the dark sector drops below the mass of the lightest state, all the energy density of the dark sector is converted into pure dark radiation. The massless $\omega$ particles will therefore play a similar role as the SM photon with respect to the dark thermal-bath and inherits from the various degrees of freedom of the dark-sector energy density. The effect of radiation energy-density $\rho_\text{rad}$ with respect to the Hubble expansion rate can be written in terms of the effective number of relativistic species $N_\text{eff}$:
\begin{equation}
    N_\text{eff}\equiv \dfrac{8}{7}\left( \dfrac{11}{4} \right)^{4/3} \left( \dfrac{\rho_\text{rad}-\rho_\gamma}{\rho_\gamma} \right)~,
\end{equation}
where $\rho_\gamma$ is the photon energy density. The SM expected value $N_\text{eff}^\text{SM}=3.046$~\cite{Mangano:2005cc,deSalas:2016ztq} differs slightly from the naïve estimate $N_\text{eff}=3$, corresponding to 3 left-handed neutrino species, due to non-instantaneous decoupling, neutrino oscillations in the plasma and finite temperature effects~\cite{Mangano:2005cc,deSalas:2016ztq,Dolgov:1997mb,PhysRevD.26.2694}. In our setup deviations from the SM expected value of $N_\text{eff}$ induced by the dark-sector energy density $\rho_\omega$ can be expressed in term of $\Delta N_\text{eff}$ as
\begin{equation}
    \rho_\text{rad}= \left( 1+ \dfrac{7}{8}\left( \dfrac{4}{11} \right)^{4/3} \left(N_\text{eff}^\text{SM}+\Delta N_\text{eff}\right) \right) \rho_\gamma ~,
\end{equation}
The dark-sector total entropy $S_\omega$ can be parametrized in term of the dark-sector temperature $T_\omega$ and the effective fermionic degrees of freedom $g^{\omega}_{\star}(T_\omega)$~\footnote{Defined in such a way for convenience as $g^{\omega}_{\star}=2$ corresponds to the $\omega$ internal degrees of freedom} as
\begin{equation}
    S_\omega=\dfrac{2\pi^2}{45} g^{\omega}_{\star}(T_\omega)T_\omega^3 a^3 ~.
\end{equation}
where $a$ is the scale factor. As detailed in Appendix~\ref{sec:appendix_deltaNeff}, by using entropy conservation arguments, the value of $\Delta N_\text{eff}$ evaluated at a temperature smaller than the electron mass $T\ll m_e$ can be expressed as  
\begin{equation}
    \Delta N_\text{eff} =\left( \dfrac{43}{4} \right)^{4/3} \left( \dfrac{g^{\omega}_{\star}(T_{\text{FO}})}{2} \right)^{4/3}  \left(  \dfrac{1}{g^\text{SM}_{\star}(T_\text{FO})}   \right)^{4/3} ~,
\end{equation}
where $g^\text{SM}_{\star}(T)$ is the usual SM-photon effective degrees of freedom. The effective number of fermionic relativistic species of the dark-sector thermal bath can be expressed as~\cite{Blennow:2012de}:
\begin{equation}
    g^{\omega}_{\star}(T)=2+\dfrac{8}{7}\sum_i \dfrac{45 g_i}{4\pi^4} z_i^4 \int_{1}^\infty \dfrac{y\sqrt{y^2-1}}{e^{yz_i}\pm1} \dfrac{4y^2-1}{3y} \diff y ~,
\end{equation}
where $i$ denotes a massive species (fermionic or bosonic) of the dark thermal bath with $z_i\equiv m_i/T$, $g_i$ denotes its internal degrees of freedom and the "+" ("-") sign in the denominator applies for fermions (bosons). As the SM effective degrees of freedom decreases sharply when the temperature drops below $T \ll T_\text{QCD}\simeq 180~\text{MeV}$ and reaches $g^\text{SM}_{\star}(T<100~\text{MeV}) \lesssim 20$, a sizable contribution $\Delta N_\text{eff}\gtrsim 0.5$ is expected and potentially can reach $\Delta N_\text{eff}=1$ if $\omega$ is the only remaining relativistic species in the dark thermal bath and the freeze-out occurs at a temperature close to the SM neutrino decoupling $T_{\nu}^\text{dec} \lesssim T \ll T_\text{QCD}$ as $\omega$ would behave as a fourth SM-like neutrino species. 
In  Appendix~\ref{sec:appendix_deltaNeff} we reported the expected values for $\Delta N_\text{eff}$ given a specific relativistic content in our model at the freeze-out temperature for illustration.

\begin{figure}[t!]
    \centering
    \includegraphics[width=0.48\linewidth]{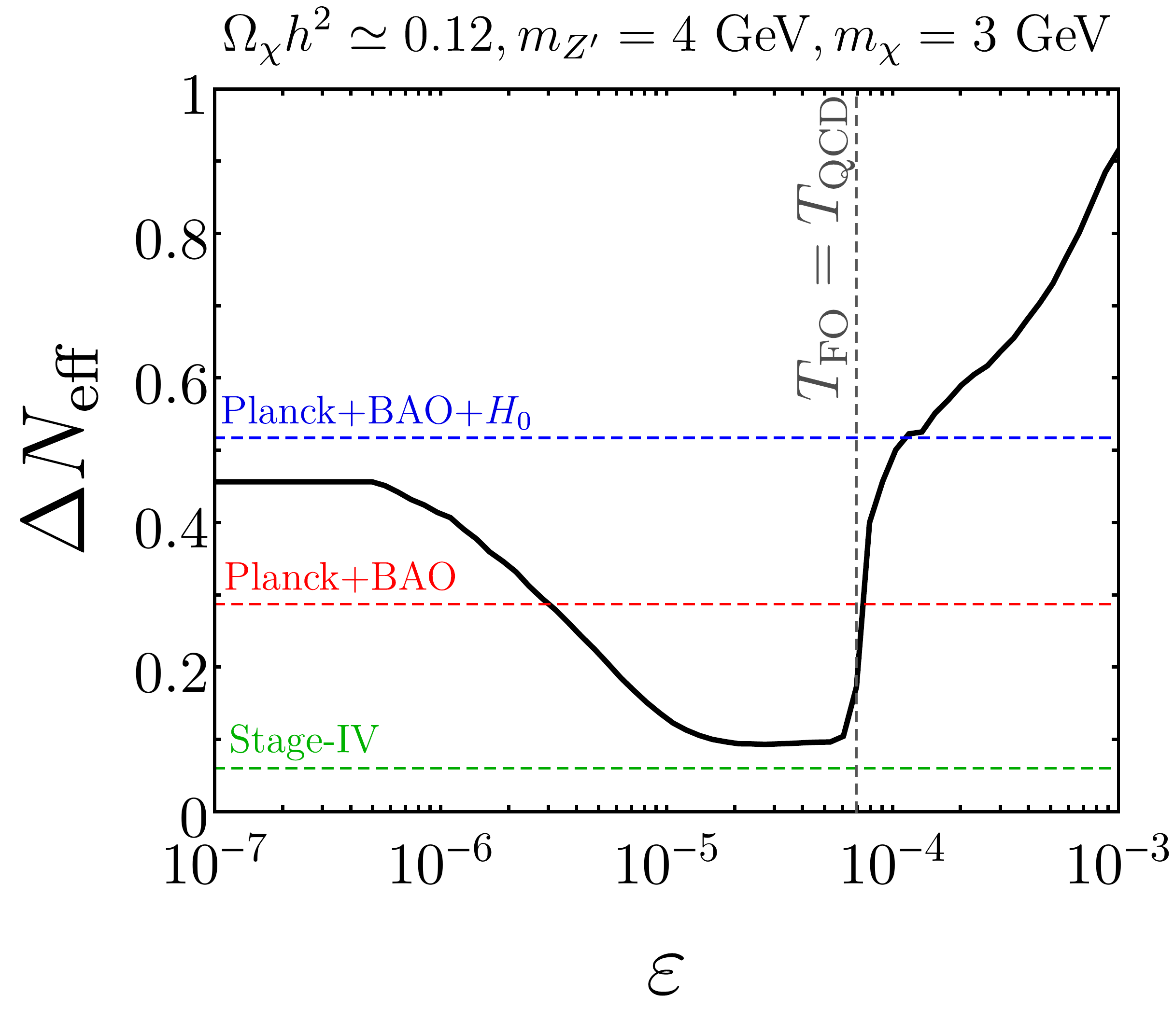}
        \includegraphics[width=0.48\linewidth]{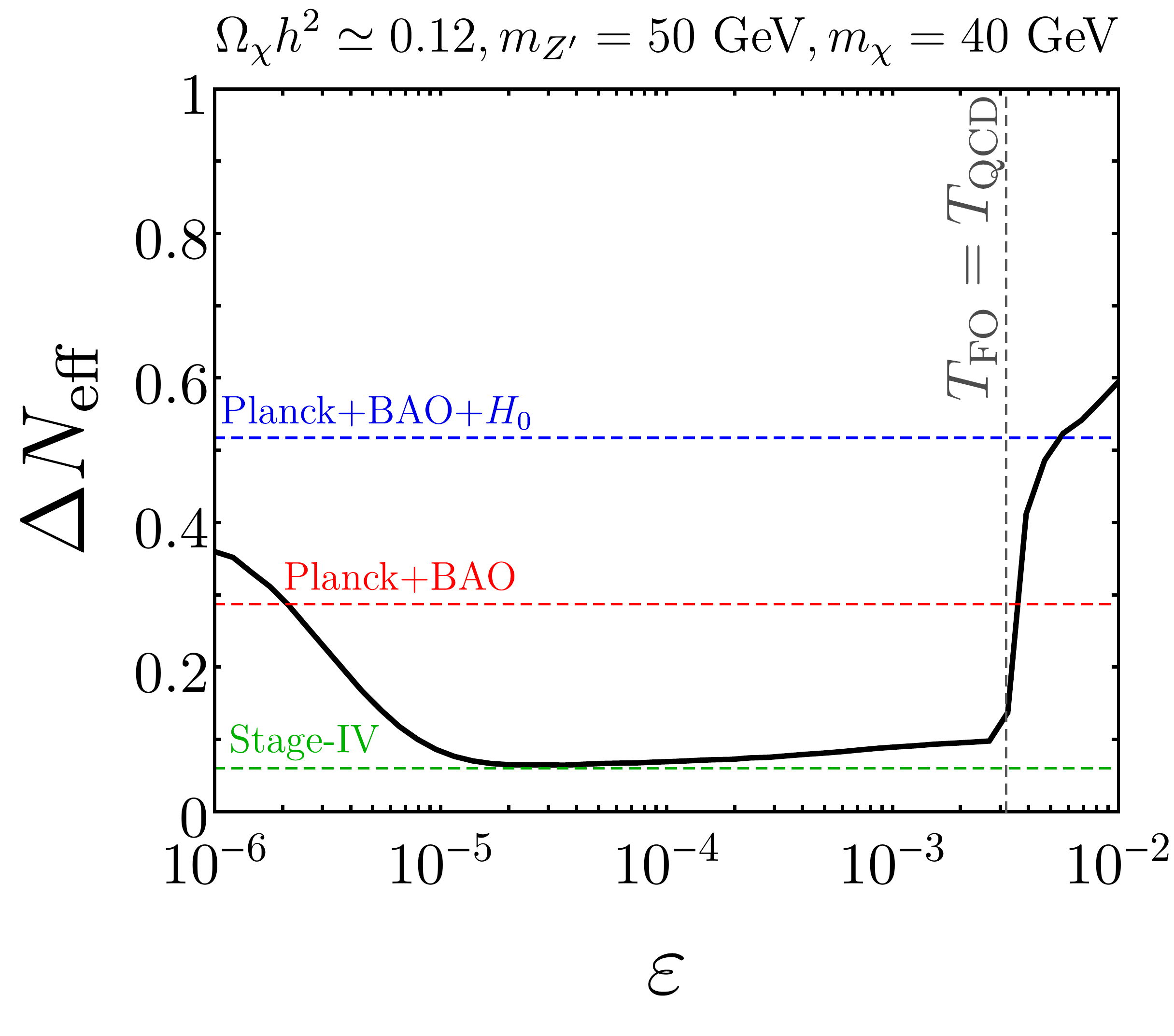}
    \caption{Expected contribution to $\Delta N_\text{eff}$ in our model for given parameters that allows to achieve the correct DM relic density. Constraints from the Planck collaboration including BAO is shown in dashed red, constraints taking into account the Hubble tension is shown in dashed blue and sensitivity prospects for the next generation of CMB experiments in shown in dashed green. Details regarding these constraints are detailed in Sec.~\ref{sec:DeltaNeff}}
    \label{fig:DeltaNeff}
\end{figure}

The precise value of $N_\text{eff}$ has important consequences on the thermal history of the universe and can be constrained by several observables. At temperatures $T\sim\text{MeV}$, it can affect the expansion rate significantly and perturb the formation of light elements during BBN. Measurements of $N_\text{eff}$ based on the standard BBN history was estimated to be rather consistent with the SM~\cite{Cyburt:2015mya} and in a recent work, it has been estimated to be~\cite{Pitrou:2018cgg}
\begin{equation}
    N_\text{eff}=2.88 \pm 0.27~~(68\%~\text{C.L.})~~~~(\text{BBN})~.
    \label{eq:BBN}
\end{equation}
At the time of recombination, another consequence of the deviation to the standard $ N_\text{eff}^\text{SM}$ is to affect the CMB power spectrum which has precisely been measured by the Planck collaboration using a combination of TT+TE+EE+lowE+lensing+BAO~\cite{Akrami:2018vks,Aghanim:2018eyx}:
\begin{equation}
    N_\text{eff}=2.99^{+0.34}_{-0.33}~~(95\%~\text{C.L.})~~~~(\text{Planck+BAO})~.
        \label{eq:Planck}
\end{equation}
In spite of a rather consistent between BBN and CMB measurements, tensions of the order of around $4-6 \sigma$~\cite{Wong:2019kwg,Verde:2019ivm} have recently been reported between measurements of the local value of the Hubble constant $H_0$~\cite{Bernal:2016gxb,Riess:2016jrr,Riess_2018,Riess:2019cxk} and the value inferred from the CMB anisotropy map by the Planck collaboration~\cite{Aghanim:2018eyx} whose most recent combined analysis has shown that a non-negligible contribution to $\Delta N_\text{eff} \sim 0.2-0.5$ could reduce the existing tension to a $\sim 3 \sigma$ level and give an estimation of the combined result
\begin{equation}
    N_\text{eff}=3.27\pm0.15~~(68\%~\text{C.L.})~~~~(\text{Planck+BAO}+H_0)~.
        \label{eq:Planck+H0}
\end{equation}
Such tensions might reveal a crisis within the $\Lambda$CDM standard cosmological model and several groups have been addressing this issue by considering early dark energy~\cite{Poulin:2018cxd,Poulin:2018dzj}, neutrino self-interactions~\cite{Kreisch:2019yzn,Park:2019ibn}, decaying dark-matter~\cite{Bringmann:2018jpr} or by considering a non-negligible contribution to $\Delta N_\text{eff} \sim 0.2-0.5$~\cite{DEramo:2018vss,Escudero:2019gzq,Mortsell:2018mfj,Vagnozzi:2019ezj}. More recently, solution to the Hubble tension have been investigated in the context of the seesaw mechanism~\cite{Escudero:2019gvw}. This class of solution alleviating the tension considering $\Delta N_\text{eff} \neq 0$ by introducing new physics in the neutrino sector is perhaps one of the simplest and the most natural which is also present in our model as sizable values of $\Delta N_\text{eff}$ are expected. \\In Fig.~\ref{fig:DeltaNeff}, we represented numerical estimates of $\Delta N_\text{eff}$ in the instantaneous freeze-out approximation using the condition of Eq.~(\ref{eq:TFO}) to determine the dark-sector freeze-out temperature in the vector mediator case and we considered values for the gauge coupling $g_{Z^\prime}$ and masses $(m_\chi, m_{Z^\prime})$ that allow to achieve the correct dark matter density $\Omega_\chi h^2\simeq 0.12$ depicted in Fig.~\ref{fig:gauge_coupling_relic_density}. In Fig.~\ref{fig:DeltaNeff}, we represented the $2\sigma$ upper bounds from the Planck analyses whose results are shown in Eq.~(\ref{eq:Planck}) and Eq.~(\ref{eq:Planck+H0}) as well as limits from the upcoming Stage-IV CMB experiment which is expected to reach a precision on the determination of $N_\text{eff}$ of around $\sim 0.03$~\cite{Abazajian:2016yjj} making a precision measurement of $N_\text{eff}$ a promising and interesting probe of our model~\footnote{The upper bound from BBN of Eq.~(\ref{eq:BBN}) is not represented as it is very similar to the value derived by the Planck collaboration.}.
\\As one could observe in Fig.~\ref{fig:DeltaNeff}, any value of the kinetic mixing $\varepsilon$ large enough to trigger a dark-sector freeze-out for a temperature smaller than the QCD scale lead to a large value of $\Delta N_\text{eff}>0.5$, in tension with Planck and BBN constraints, therefore we consider the corresponding large values of $\varepsilon$ to be excluded. On the other hand, when decreasing $\varepsilon$ the freeze-out occurs earlier and having only few remaining relativistic states lead to low values of $\Delta N_\text{eff}\sim0.05-0.1$. Small values of $\varepsilon$ represented in Fig.~\ref{fig:DeltaNeff}, correspond to the regime where the dark sector is produced by $Z^\prime$ inverse-decay, achieve thermalization with the SM bath for $T\sim m_{Z^\prime}$, and decouples when the temperature drops below the $Z^\prime$ mass, leaving relativistic states in the thermal bath and leading to larger contribution to $\Delta N_\text{eff}\sim0.1-0.5$. 
A more precise determination of $\Delta N_\text{eff}$ in this model is beyond the scope of this work and would require a more detailed analysis. As non-instantaneous decoupling and thermal effects have been shown to contribute at the order of $\sim0.05$~\cite{deSalas:2016ztq}, our $\Delta N_\text{eff}$ predictions are expected to be correct at the same order of magnitude. 
Changing the values of the masses chosen in Fig.~\ref{fig:DeltaNeff} would tend to prefer large values of $\Delta N_\text{eff}\sim 0.3-0.5$ for lower masses $m_\chi,m_{Z^\prime}\sim\text{GeV} $ and smaller values $\Delta N_\text{eff}\sim 0.1-0.3$ for larger masses $m_\chi,m_{Z^\prime}\sim10-100~\text{GeV}$ while still achieving the correct relic density. Typically the heavy neutrino states $N_{4,5}$ have to be heavier than $m_\chi,m_{Z^\prime}$ and become non-relativistic at the time of freeze-out as they would induce a value of  $\Delta N_\text{eff}$ larger than the present constraints.

\subsection{Thermalization via scalar and neutrino portal}

In this section we describe how to achieve thermalization and freeze-out in the regime where interactions between the dark sector and the SM particles are mainly due to either the scalar mixing angle $\sin \alpha$ or the neutrino Yukawa coupling $y_\nu$. We first start by discussing the scalar mixing case. Similarly to the kinetic mixing case, the dark sector can be produced from the SM thermal bath by inverse-decay induced by mixing of the mediator $\phi$  with the SM Higgs, whose decay width to a pair of SM fermions $\psi$ is given by
\begin{equation}
    \Gamma_{\phi \rightarrow \bar{\psi}\psi}=c_\psi \frac{s_{\alpha }^2  m_\phi m_{\psi }^2 }{8 \pi  v_h^2}\left(1-\frac{4 m_{\psi }^2}{m_\psi^2}\right)^{3/2}~,
\end{equation}
where $c_\psi$ is a color factor. The main difference to the vector mediator case is that the partial width is proportional to the SM fermion masses $m_\psi$, therefore the width is expected to be very narrow in the case where $m_\phi$ is rather light, below the GeV scale. For this reason, inverse decays of $\phi$ is typically less efficient compared to the vector mediator case but still able to make the dark sector thermalize. Applying the condition of Eq.~(\ref{eq:condition_Zprime_thermalization}) to the scalar case gives
\begin{equation}
   \sum_\psi  \Gamma_{\phi \to \psi\bar \psi}    \gtrsim \left. \dfrac{H(z)z}{Y_{Z^\prime}^{\rm (eq)}(z) } \frac{ K_2(z) }{  K_1(z) } \right|_{z=1} ~,
  \label{eq:condition_Phi_thermalization}
\end{equation}
with $z\equiv m_\phi/T$. This equation provides a lower bound on $\sin \alpha$ for which the scalar $\phi$ is produced and thermalizes with the SM bath at $z=1$. For a typical mass $m_\phi \sim 1-100~\text{GeV}$ the bound on $\alpha$ is 
\begin{equation}
\sin^2 \alpha \gtrsim 5 \times 10^{-9}~,
\end{equation}
and is almost constant in this mass range. Another efficient way of producing the dark sector is to consider $\bar t t$ annihilation into a DM pair, as the top-quark Yukawa is large, in the case where the DM Yukawa is large as well, this process can be very efficient. As detailed in Sec.~\ref{sec:producing_dark_sector_via_2to2} the production of a DM pair from SM particle annihilation can be estimated by relating the evolution DM yield $Y_\chi$ to the dimensionless rate of the $\bar t t \to \bar \chi \chi$ process via the Boltzmann equation
\begin{equation}
\dfrac{\diff Y_\chi}{\diff z_t}=R_{\bar t t \to \bar \chi \chi}(z_t)\simeq y_\chi^2 \left( \dfrac{\sin^2 \alpha}{2 \times 10^{-10}} \right)~,
\label{eq:collision_term_ttbartochichi}
\end{equation}
where we have defined $z_t\equiv m_t/T$ with $m_t\simeq 173~\text{GeV}$ being the top-quark mass. Using the condition $R(z_t=1)\gtrsim 1$ derived in Sec.~\ref{sec:producing_dark_sector_via_2to2}, the DM yield will reach its equilibrium value around $z_t\sim 1$ for $\sin^2\alpha \gtrsim 2 \times 10^{-10}$ which gives a relaxed lower bound on $\sin \alpha$ for a large Yukawa $y_\chi \sim 1$. As in the vector mediator case, interaction rates in the dark sector induced by a gauge coupling $g_{Z^\prime}\sim 10^{-2}-10^{-1}$ are expected to be efficient enough to allow for the production of the entire dark sector in one of its species reaches equilibrium with the SM. However since the coupling between the dark scalar $\psi$ and SM fermions is proportional to their masses, the scattering rate is reduced and therefore the freeze-out is expected to occur earlier. As the massless field $\omega$ do not couple to $\phi$, in this case thermal equilibrium between both sectors depends on the scattering rate between DM and SM particles~\footnote{As discussed in the previous section, if heavy neutrinos $N_i$ are still relativistic and thermalized in the dark sector at the freeze-out temperature, the value of $\Delta N_\text{eff}$ is expected to be large and excluded.}. To estimate the freeze-out condition we can write the time evolution of the dark sector temperature and express it as a function of the momentum relaxation rate $\gamma_{\chi \psi}$, which is given in the non-relativistic limit by 
\begin{equation}
    \dfrac{\diff T_\chi}{\diff t}+2HT_\chi\simeq -2\gamma_{\chi \psi}(T)( T_\chi- T)~,
    \label{eq:darktemp}
\end{equation}
where $\gamma_{\chi \psi}(T)$ is the momentum relaxation rate whose definition can be found in~\cite{Choi:2017zww} and is given by
\begin{equation}
    \gamma_{\chi \psi} (T)=c_\psi \sin^2 \alpha \frac{31 \pi  y_{\chi }^2  m_{\psi }^2 T^6}{756 v_h^2 m_{\chi }} \left( \dfrac{1}{m_h^2}-\dfrac{1}{m_\phi^2}\right)^2~,
\end{equation}
Comparing this expression to the Hubble expansion rate at a temperature of the order of the QCD scale $T\sim T_\text{QCD}\simeq 180~\text{MeV}$ gives
\begin{equation}
\dfrac{\gamma_{\chi \psi}(T_\text{QCD})}{H(T_\text{QCD})}\simeq 4  \sin^2 \alpha \left( \dfrac{y_{\chi}}{10^{-2}} \right)^2 \left( \dfrac{10~\text{GeV}}{m_{\phi}} \right)^4 \left( \dfrac{10~\text{GeV}}{m_{\chi}} \right) ~,
\end{equation}
where we took $\psi$ to be a GeV scale SM quark. As large mixing angles are needed for this ratio to be $\mathcal{O}(1)$, in most of the parameter space the dark-sector freeze-out occurs before the QCD phase transition, implying expected values for $\Delta N_\text{eff} \in [0.05,0.5 ]$, i.e. in the same range than in the allowed parameter space for the vector mediator case.\\
The possibility to produce a thermalized dark sector from the interactions between SM-like and heavy neutrinos 
has been studied in~\cite{Besak:2012qm, Garbrecht:2013urw,Ghisoiu:2014ena}
where it has been shown that for $y_\nu \gtrsim10^{-7}$  
 thermalization can be achieved between the hidden-sector and the SM before the electroweak phase transition.
However, in this case, as the same Yukawa coupling is responsible for both the production and freeze-out of the dark sector, if the the dark-sector freezes-out when the heavy neutrinos $N_{4,5}$ are still relativistic, as discussed in the previous section, contribution to the relativistic degrees-of-freedom might exceed the value currently allowed by BBN or CMB measurements, making this option difficult to reconcile with experiments~\footnote{This might not be the case if some hierarchy is present among the heavy-neutrinos Yukawa coupling but would require a dedicated analysis, which is beyond the scope of the paper.}.

\section{Constraints}

\label{sec:pheno}

\subsection{Constraints on the mediators}

The mixing between the scalar $\phi$ with the Higgs induces a deviation from the SM-expected couplings of the Higgs to gauge bosons $\kappa_V$ and fermions $\kappa_f$ by a factor $\kappa_F=\kappa_V= \cos\alpha$ which is constrained to be $\sin\alpha<0.2-0.3$ for $m_\phi$ between 200-800 GeV~\cite{Robens:2016xkb}. If the scalar is much lighter than the Higgs, for instance in the region $1<m_\phi <10$ GeV, the constraints on the mixing range from $\sin\alpha>10^{-3}-10^{-1}$~\cite{Clarke:2013aya}.
If $m_\phi<m_h/2$ the Higgs can decay into a pair of scalars which  contributes to the invisible width, constrained to be $ \Gamma_{h\rightarrow \text{inv}}/\Gamma_h<24 \% $~\cite{Tanabashi:2018oca} with a total SM-expected width of $\Gamma_{h}=4.12~\text{MeV}$~\cite{Aad:2015xua}. The same argument applies to the decay of the Higgs into a pair of hidden-sector fermions. Moreover if the decay $B^+ \to K^+ \phi$ is kinematically allowed, for $m_\phi \lesssim 5~\text{GeV}$, the BaBar limit $\text{Br}(B^+ \to K^+ \bar \nu \nu)<1.6 \times 10^{-5}$~\cite{Lees:2013kla} constrains the mixing angle of the order of $\sin^2 \alpha \gtrsim 5 \times 10^{-6}$. We used the bound as derived in Ref.~\cite{Krnjaic:2015mbs}.

\begin{figure}[h!]
    \centering
    \includegraphics[width=0.46\linewidth]{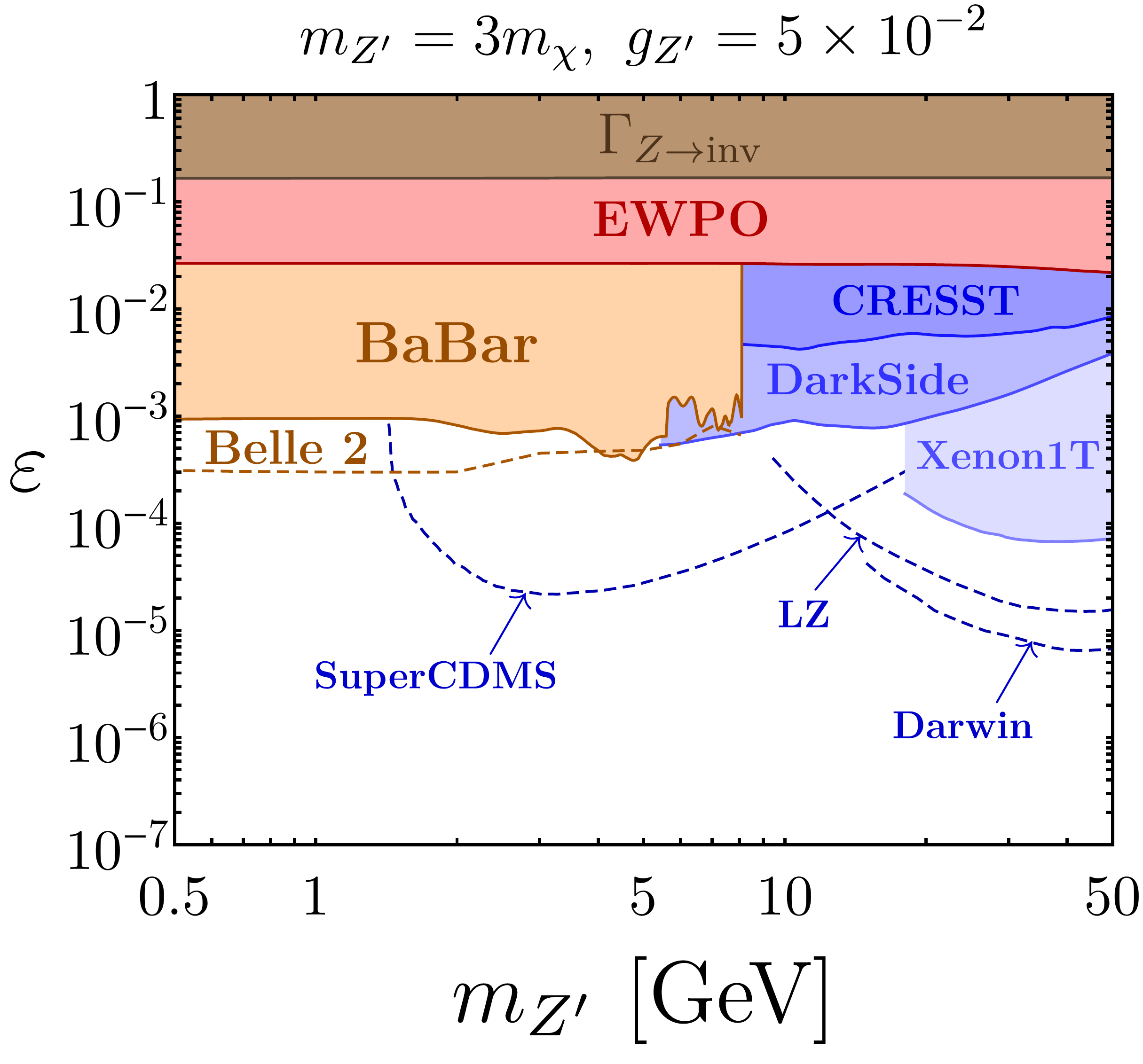}
        \includegraphics[width=0.49\linewidth]{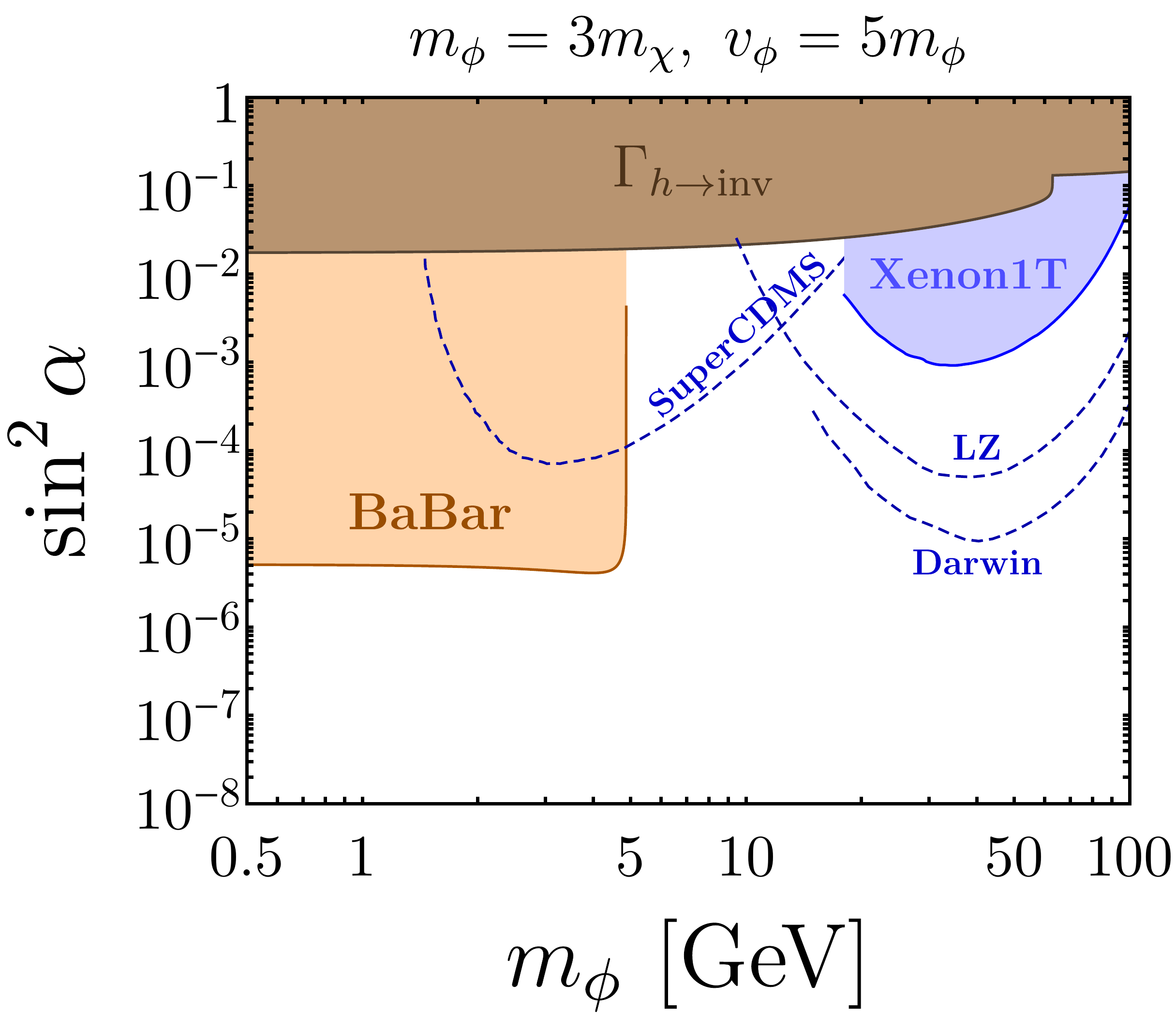}
    \caption{Constraints on the mixings $\varepsilon$ between $Z^\prime$ and SM hypercharge and $\alpha$ between the new scalar $\phi$ and the Higgs from Invible decays of $Z$ and $h$, BaBar, ElectroWeak Precision Observables (EWPO) and current direct detection experiments (CRESST, DarkSide, Xenon1T) as well as
    sensitivity estimation from Belle 2 and future direct detection experiments (SuperCDMS, LZ, Darwin). All the constraints represented in this figure are detailed in Sec.~\ref{sec:pheno}.}
    \label{fig:constraints_epsilon_VS_mzp}
\end{figure}

The $Z-Z'$ mixing affects the $Z$-boson width which has been measured~\cite{Tanabashi:2018oca} as $\Gamma_Z= 2.4952 \pm 0.0023~\text{GeV}$ with an invisible contribution $\Gamma_Z= 499.0 \pm 1.5~\text{MeV}$. Since the interesting part of the parameter space, the $Z^\prime$ decays invisibly, it can be constrained by mono-photon searches in $e^+e^-$ annihilations with BaBar~\cite{Lees:2017lec} and from various ElectroWeak Precision Observables (EWPO)~\cite{Curtin:2014cca}. According to Ref.~\cite{Kou:2018nap}, Belle 2 should be able to improve sligthly the sensitivity of BaBar. A summary of these  constraints is depicted in Fig.~\ref{fig:constraints_epsilon_VS_mzp}.

\subsection{Direct detection}
\subsubsection{Vector mediators portal}
As detailed in Sec.~\ref{sec:thegaugesector}, our $Z^\prime$ couples only to the SM EM current, therefore $Z^\prime$ mediated DM-nucleus scatterings would only trigger Spin-Independent (SI) cross section. However since our DM candidate can also interact with the $Z$ boson, this mediator contributes to both Spin-Independent and Spin-Dependent (SD) cross sections as axial-vector interactions are also present in our model. The low-energy relevant DM-quark effective operator relevant for SI and SD cross sections are
\begin{equation}
    \mathcal{O}_{\chi q}^\text{SI}=C_{\chi q}^\text{SI} \Big( \bar{\chi} \gamma_\mu \chi \Big) \Big( \bar{q} \gamma^\mu q \Big)~,\quad~ \mathcal{O}_{\chi q}^\text{SD}=C_{\chi q}^\text{SD} \Big( \bar{\chi} \gamma_\mu \gamma_5 \chi \Big) \Big( \bar{q} \gamma^\mu \gamma_5 q \Big)~,
\end{equation}
with
\begin{equation}
  C_{\chi q}^\text{SI} = \varepsilon g_{Z^\prime} V_\chi\left(\dfrac{ q_L^Z+q_R^Z}{2m_{Z}^2} - \dfrac{  Q_q  e}{m_{Z^\prime}^2}  \right) ~,\quad~ C_{\chi q}^\text{SD} = \varepsilon t_W   g_{Z^\prime} A_\chi
  \left(\dfrac{ q_L^Z-q_R^Z}{2m_{Z}^2}  \right) ~,
\end{equation}
where $Q_q$ is the electric charge of the SM-quark $q$. $q_{L,R}^Z$ are respectively the couplings from left and right-handed quarks to the $Z$ boson. $V_\chi$ and $A_\chi$ are the vector and axial-vector DM-$Z^\prime$ coupling defined in Sec.~\ref{sec:modelthedarksector}. The corresponding DM-nucleon effective operator is given by
\begin{equation}
        \mathcal{O}_{\chi N}^\text{SI}=C_{\chi N}^\text{SI} \Big( \bar{\chi} \gamma_\mu \chi \Big) \Big( \bar{N} \gamma^\mu N \Big)~,\quad ~\mathcal{O}_{\chi N}^\text{SD}=C_{\chi N}^\text{SD} \Big( \bar{\chi} \gamma_\mu \gamma_5 \chi \Big) \Big( \bar{N} \gamma^\mu \gamma_5 N \Big)~,
\end{equation}
where $N=p,n$ with $p$ and $n$ denote proton and neutron. The SI Wilson coefficients are $C_{\chi p}^\text{SI}=2C_{\chi u}^\text{SI}+C_{\chi d}^\text{SI}$ and $C_{\chi n}^\text{SI}=C_{\chi u}^\text{SI}+2C_{\chi d}^\text{SI}$. The SD coefficients can be expressed as a sum over light quarks as $C_{\chi N}^\text{SD}=\sum_q C_{\chi q}^\text{SD} \Delta_q^N$ with the coefficients $\Delta_q^N$ are given in Ref.~\cite{DelNobile:2013sia}. The total averaged\footnote{As  bounds on the DM-nucleon cross section are derived by experimental collaborations under the assumption of isospin symmetry, the quantity that could be compared to such limits has to be averaged over protons and neutrons.} DM-nucleon SI and SD cross sections are given by~\cite{Arcadi:2017atc,Arcadi:2015nea,Pierre:2016wgb}
\begin{equation}
   \hat \sigma_{N}^\text{SI}=\dfrac{\mu_{\chi N}^2}{\pi} \left[ C_{\chi p}^\text{SI} \dfrac{Z}{A}+ C_{\chi n}^\text{SI} \left( 1-\dfrac{Z}{A}\right)\right]^2~,\quad~ \hat\sigma_{N}^\text{SD}=\dfrac{3\mu_{\chi N}^2}{\pi} \dfrac{\left[ C_{\chi p}^\text{SI} S_p^A+  C_{\chi n}^\text{SI}  S_n^A  \right]^2}{\left[ S_p^A+ S_n^A  \right]^2}~,
\end{equation}
where $\mu_{\chi N}$ is the DM-nucleon reduced mass, $Z$ and $A$ are respectively the number of protons and nucleons present in an atom of detector material~\footnote{For a more precise estimate we would have to average this formula over isotope relative abundance, in practise the improvement only marginal.}, $S_{N}^A$ correspond to the contribution of the nucleon $N$ to the spin of a nucleus with $A$ nucleon. Typically for Xenon-based detectors $S_{n}^A \gg S_{p}^A$ therefore cross section with neutrons can be used as reference. \\
Regarding SI cross section, as the $Z^\prime$ couple only to the EM current in the SM sector, the contribution from this mediator to the neutron cross-section is vanishing. Therefore in the limit $m_{Z^\prime} \ll m_Z$ the proton cross section is expected to be larger and is given by
\begin{equation}
    \sigma_p^\text{SI}=\dfrac{m_p^2 m_\chi^2}{\pi (m_p+ m_\chi)^2 m_{Z^\prime}^4} e^2 g_{Z^\prime}^2 V_\chi^2 \varepsilon^2\simeq 2 \times 10^{-46}~\text{cm}^2~\left(\dfrac{\varepsilon}{10^{-5}} \right)^2 \left(\dfrac{g_{Z^\prime}}{10^{-2}} \right)^2 \left( \dfrac{10~\text{GeV}}{m_{Z^\prime}} \right)^4~,
\end{equation}
in the limit $m_\chi \gg m_p$, while the neutron cross section is 5 order of magnitude smaller for the same choice of parameters. In the case of SD interactions, we get the following numerical approximation for neutrons
\begin{equation}
    \sigma_n^\text{SD}\simeq 1.6 \times 10^{-52}~\text{cm}^2~\left(\dfrac{\varepsilon}{10^{-5}} \right)^2 \left(\dfrac{g_{Z^\prime}}{10^{-2}} \right)^2~.
\end{equation}
These expressions are in agreement with numerical computations obtained using the public code micrOMEGAs~\cite{Belanger:2013oya}.

\subsubsection{Scalar mediators}
In the case where the DM-nucleon scattering is mediated by scalars $h$ and $\phi$, only SI interactions are generated at the nuclear scale, whose effective operator can be written as
\begin{equation}
    \mathcal{O}_{\chi q}^\text{SI}=C_{\chi q}^\text{SI}\bar{\chi} \chi \bar{q} q\quad\text{with}\quad 
  C_{\chi q}^\text{SI}=\dfrac{\sin (2 \alpha) m_q m_\chi}{2v_h v_\phi} \left(\dfrac{1}{m_\phi^2}-\dfrac{1}{m_h^2} \right)~,
 \end{equation} 
giving rise to the following DM-nucleon effective operator
\begin{equation}
    \mathcal{O}_{\chi N}^\text{SI} = C_{\chi N }^\text{SI} \bar{\chi} \chi \bar{N} N ~,
\end{equation}
and corresponding averaged DM-nucleon SI cross section:
\begin{equation}
   \hat \sigma_{N}^\text{SI}=\dfrac{\mu_{\chi N}^2}{\pi} \left[ C_{\chi p}^\text{SI} \dfrac{Z}{A}+ C_{\chi n}^\text{SI} \left( 1-\dfrac{Z}{A}\right)\right]^2~,
\end{equation}
where 
\begin{equation}
    C_{\chi N}^\text{SI}=\dfrac{\sin (2\alpha) m_N m_\chi}{2v_h v_\phi} \left(\dfrac{1}{m_\phi^2}-\dfrac{1}{m_h^2} \right) \Big(f_{Tu}^{(N)}+f_{Td}^{(N)}+f_{Ts}^{(N)}+\dfrac{6}{27}f_{TG}^{(N)}\Big)~,
\end{equation}
where $f_{Tq}^{(N)}\equiv \langle N | m_q \bar q q | N \rangle/m_N$ and $f_{TG}^{(N)}\equiv 1 - \sum_q f_{Tq}^{(N)}$ whose numerical values can be find in Ref.~\cite{DelNobile:2013sia}. In the limit $m_\chi \gg m_p$ and $m_\phi \ll m_h$ we get the following numerical estimate for the proton cross-section\footnote{The neutron cross-section is almost identical.}:
\begin{equation}
    \sigma_p^\text{SI} \simeq 7 \times 10^{-46}~\text{cm}^2~\left( \dfrac{\sin \alpha}{10^{-3}} \right)^2     \left( \dfrac{m_{\chi}}{10~\text{GeV}} \right)^2     \left( \dfrac{10~\text{GeV}}{m_{\phi}} \right)^4    \left( \dfrac{50~\text{GeV}}{v_{\phi}} \right)^2~.
\end{equation}

\subsubsection{Present and future bounds}
The most constraining direct detection experiment to this day regarding SI interactions is Xenon1T~\cite{Aprile:2018dbl} which constrains DM masses above $m_\chi \gtrsim 5~\text{GeV}$ to be typically $\sigma_{\chi N}^\text{SI} \lesssim 10^{-47}~\text{cm}^2$ at $m_{\chi}\sim 50~\text{GeV}$. The next generation of xenon-based experiments such as LZ~\cite{Akerib:2018dfk} or Darwin~\cite{Aalbers:2016jon} are expected to improve the current sensitivity by more than 1 or 2 orders of magnitude and might reach the neutrino floor~\cite{Billard:2013qya}. At lower masses $m_\chi \sim 1-10~\text{GeV}$, bounds from the most sensitive experiments such as DarkSide~\cite{Agnes:2018ves} and  CRESTIII~\cite{Abdelhameed:2019hmk} which should be extended in the future at lower masses $m_\chi \gtrsim 500~\text{MeV}$ by the cryogenic detector SuperCDMS~\cite{Agnese:2016cpb}.
Concerning SD interactions, current experiments are not as sensitive, the most stringent bound in derived by the PICO-60 bubble chamber~\cite{Amole:2017dex} which constrains $\sigma_{\chi N}^\text{SI} \lesssim 10^{-41}~\text{cm}^2$ for masses $m_{\chi}\sim 30~\text{GeV}$. The current bounds and sensitivity prospects described in this section are depicted in Fig.~\ref{fig:constraints_epsilon_VS_mzp}.\\
Essentially, in the kinetic-mixing case, constraints from monophoton searches with the BaBar experiment are the most stringent for masses $m_\chi \lesssim 10~\text{GeV}$, threshold from which direct detection constraints starts to dominate, excluding values $\varepsilon \gtrsim 10^{-3}$ and potentially until $\varepsilon \gtrsim 10^{-4}-10^{-5}$ in the future~\footnote{SD constraints are not represented as they are less competitive than SI in our model.}. However, in the scalar mixing case, direct detection experiments are less constraining, excluding $\sin^2 \alpha \gtrsim 10^{-3}$ at masses of the order of the GeV scale, while kaon-decay constraints from the BaBar experiment impose $\sin^2 \alpha \lesssim 10^{-5}$ for masses $m_\chi \lesssim 5~\text{GeV}$. As direct-detection signatures are expected to be quite different in both kinetic mixing and scalar mixing case, they offer a complementary way of probing the model.

\subsection{Indirect detection}
In our model, the DM states annihilate predominantly into dark-sector particles, making indirect detection signatures essentially absent in this model, in the part of the parameter space which is not already excluded by other experiments. However, in the particular case where $2 m_\chi \sim m_{Z^\prime}$ the gauge coupling required to achieve the correct relic abundance is significantly reduced as the cross section in enhanced by the resonance, as shown in Fig.~\ref{fig:gauge_coupling_relic_density}. For masses typically below $\sim 10~\text{GeV}$, constraints from energy-injection of dark matter annihilation during the dark ages into charged particles, place a bound on the cross section of the order of $\langle \sigma v \rangle_{\bar \chi \chi \to e^+e^-} \lesssim 10^{-27}\text{cm}^{3}\text{s}$~\cite{Slatyer:2015jla} which is below the typical value $\langle \sigma v \rangle_{\bar \chi \chi \to \omega \omega} \lesssim 3 \times 10^{-26}\text{cm}^{3}\text{s}^{-1}$. As in our model both quantities are $s-$wave dominated and the ratio of these quantities goes as 
\begin{equation}
    \dfrac{\langle \sigma v \rangle_{\bar \chi \chi \to e^+e^-}}{\langle \sigma v \rangle_{\bar \chi \chi \to \omega \omega}}\simeq \dfrac{e^2 \varepsilon^2}{8 g_{Z^\prime}^2},
\end{equation}
for low value of the gauge coupling, the relative contribution from the electron channel becomes more important and allows to probe small values $\varepsilon\sim 10^{-4}$ precisely on the resonance and offers a clear signature for these very specific parameters.

\section{Discussion and conclusion}
\label{sec:results}

\begin{figure}[t!]
    \centering
    \includegraphics[width=0.49\linewidth]{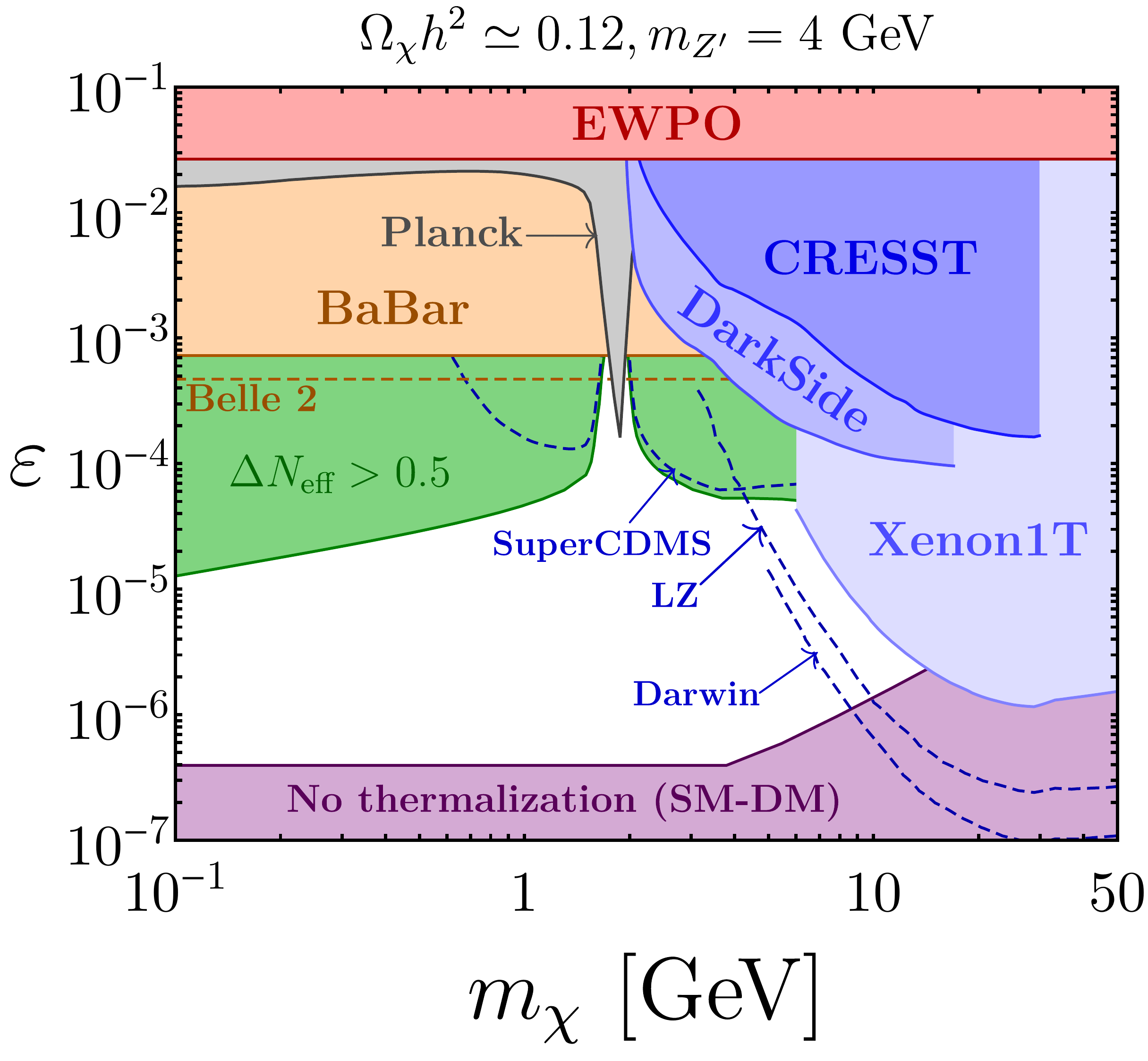}
        \includegraphics[width=0.49\linewidth]{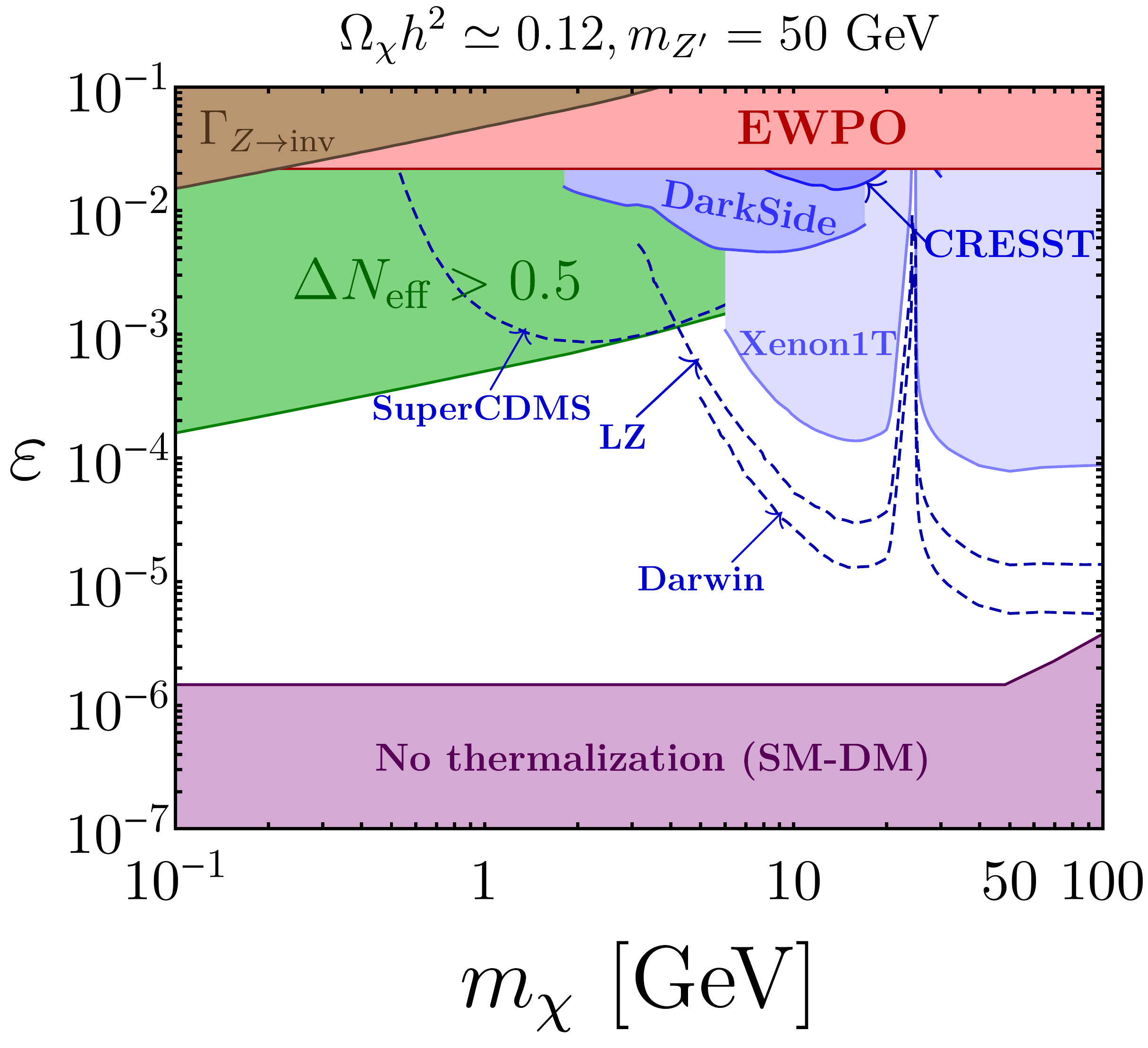}
    \caption{Constraints on the parameter space compatible with the observed dark matter abundance and neutrino masses. Present bounds are shown in solid lines and future bounds in dashed lines. Only the region in white is allowed by current experiments and predicts $\Delta N_\text{eff}\sim 0.05-0.5$. Various constraints depicted in this figure are described in the text of Sec.~\ref{sec:results}.}
    \label{fig:eps_VS_mchi}
\end{figure}

In this section we present a summary of the constraints and signatures of the model in the case where the portal between the hidden and SM sector is ensured by the kinetic mixing parameter $\varepsilon$. In Fig.~\ref{fig:eps_VS_mchi}, we represented in the plane $(m_\chi,\varepsilon)$ the various constraints detailed in this paper satisfying both observed neutrino masses and correct dark-matter relic abundance by considering numerical results depicted in Fig.~\ref{fig:gauge_coupling_relic_density} for 2 different values of $m_{Z^\prime}$. Constraints from present and future direct detection experiments (CRESST, DarkSide, Xenon1T, SuperCDMS, LZ, Darwin) are represented in blue while constraints on the mediator described in Sec.~\ref{sec:pheno} are represented in orange (Babar, Belle 2), red (EWPO) and brown ($\Gamma_{Z \to \text{inv}}$).
Future constraints are depicted in dashed lines while current constraints are in solid lines. As described in Sec.~\ref{sec:pheno}, indirect detection constraints are mostly sensitive to the pole region $2 m_\chi \sim m_{Z^\prime}$ where DM annihilations are enhanced by the $Z^\prime$ resonance and are depicted in gray (Planck). Constraints from cosmology are represented in green ($\Delta N_\text{eff}>0.5$) corresponding to the parameter space where the dark-sector freeze-out occurs after the QCD phase transition. The region in purple represents the part of the parameter space where the kinetic mixing $\varepsilon$ is basically to small to thermalize the dark sector with the SM ("No thermalization (SM-DM)")~\footnote{Notice that the requirement for the dark sector to thermalize at some point in the early universe with the SM plasma is not necessary to achieve the correct DM relic abundance in this framework. However, the detailed study of the phenomenology of the model in this regime would require a dedicated analysis, which goes beyond the scope of this work.}. In most of the parameter space, the gauge coupling is large $g_{Z^\prime} \gg \varepsilon$ and thermalization is expected to occur within the dark sector. Since the dark-sector is produced by $Z^\prime$ inverse-decays, as minimum thermalization condition we required the temperature at which the dark sector termalize with the SM to be $T = \text{max}[m_{\chi},m_{Z^\prime}]$ such that the DM thermalizes with the SM before becoming non-relativistic, in order to preserve the DM freeze-out mechanism. 
Essentially, values of the kinetic mixing $\varepsilon \sim 10^{-4}-10^{-6}$ spanning masses $m_\chi \sim 0.1-100~\text{GeV}$, are compatible with the observed dark matter relic abundance and evade all present bounds. The case of the scalar mixing is less constrained and more parameter-dependent, therefore not represented but can account simultaneously as well for both dark matter relic abundance and neutrino masses. Interestingly, as described in Sec.~\ref{sec:DeltaNeff} in all the parameter space allowed, values of $\Delta N_\text{eff}$ from $\sim 0.05$ to $0.5$ are predicted in this model making it testable by the next generation of CMB experiments while allowing to alleviate the recently established Hubble tension. Our model predicts both SI and SD DM direct detection signatures in the case of the kinetic mixing and only SI in the scalar portal case, and could be used in addition to other signatures, whose large diversity is illustrated in Fig.~\ref{fig:eps_VS_mchi}, to probe this specific model.\\

To summarize, in this paper we investigated the phenomenology of a hidden-sector model accounting both for neutrino masses and the observed dark matter relic abundance. The dark sector exhibits a local dark $U(1)'$ symmetry under which the SM fields are singlets. 
We introduced two copies of right-handed neutrinos needed to explain neutrino masses in the extended seesaw framework where only one copy is charged under the $U(1)'$ symmetry. To ensure an anomaly-free theory, additional dark chiral fermions are added to the field content.
Upon spontaneous breaking of the $U(1)'$ symmetry two of the fermions become massive whereas one state remains massless. Furthermore, the breaking of the $U(1)'$ symmetry allows for all right-handed neutrinos become massive, and gives rise to a dark $Z'$ gauge boson. The breaking up the $U(1)'$ symmetry occurs typically at the electroweak scale which sets the mass scale for the new physics particles. Due to an accidental global symmetry in the hidden sector one of the new fermions is stable and a viable DM candidate.
We have investigated the thermal history of the dark sector in the case of the kinetic mixing portal and found that due to the rather strong interactions of the dark particles~\footnote{as compared to the typical dark-sector SM interactions.} required to achieve the correct DM relic density, thermalization of the dark sector is ensured once a large population of $Z'$ is produced by inverse-decay. The scalar-mixing portal was investigated as well and was shown to be able to reach the same thermal equilibrium state without being excluded by present constraints. The massless particle $\omega$ plays the same role as the photon in the SM and inherits from the various degrees of freedom of the dark-sector thermal bath once it freezes-out from the SM plasma. As a result, the effective number of relativistic species in the universe is affected and gives a specific signature of our model, whose value predicted in the allowed parameter space should be tested by the next generation of CMB experiments and provides a way to relax the recently observed tension between early and late measurements of the Hubble constant.

\acknowledgments

The authors would like to thank Enrique Fernández-Martinez and Pedro A. N. Machado for very helpful discussions. The work of MP was supported by the Spanish Agencia Estatal de Investigaci\'{o}n through the grants FPA2015-65929-P (MINECO/FEDER, UE),  PGC2018-095161-B-I00, IFT Centro de Excelencia Severo Ochoa SEV-2016-0597, and Red Consolider MultiDark FPA2017-90566-REDC. MP would like to thank the Lawrence Berkeley National Laboratory for its hospitality during part of the realization of this work as well as the
Paris-Saclay Particle Symposium 2019 with the support of the P2I and SPU research departments and
the P2IO Laboratory of Excellence (program “Investissements d’avenir”
ANR-11-IDEX-0003-01 Paris-Saclay and ANR-10-LABX-0038), as well as the
IPhT. This project has received funding/support from the European Unions Horizon 2020 research and
innovation programme under the Marie Skodowska-Curie grant agreements Elusives ITN No. 674896
and InvisiblesPlus RISE No. 690575. J.G. is supported by the US Department of Energy under Grant Contract DE-SC0012704.

\bibliography{bibfile}{}

\appendix  

\section{Computation of $\Delta N_\text{eff}$}
\label{sec:appendix_deltaNeff}
The expression of $\Delta N_\text{eff}$ used in Sec.~\ref{sec:DeltaNeff} can be derived by entropy conservation conditions in 3 sectors: $\gamma$, $\nu$ and $\omega$ at the dark sector freeze-out temperature $T_\text{FO}$, SM neutrino decoupling ($T_\nu^\text{dec}$) and a temperature $T$ much lower than the electron mass, where $\gamma$ represents the sector composed of photons and SM particles except neutrinos such that $\text{SM}=\gamma+\nu$ and $\omega$ the dark thermal bath. Entropy conservation between the dark sector freeze-out and neutrino decoupling temperature gives 
\begin{equation}
    g^{\omega}_{\star}(T_\text{FO})T_\text{FO}^3 =g^{\omega}_{\star}(T_{\nu\text{dec}})T_\omega^3(T_{\nu\text{dec}})~,
\end{equation}
and
\begin{equation}
    g^{\text{SM}}_{\star}(T_\text{FO})T_\text{FO}^3 =g^{\text{SM}}_{\star}(T_{\nu\text{dec}})T_{\nu\text{dec}}^3~,
\end{equation}
giving the following expression for the dark sector at the neutrino decoupling temperature
\begin{equation}
    \dfrac{T_\omega^3(T_{\nu\text{dec}})}{T_{\nu\text{dec}}^3}=\dfrac{ g^{\omega}_{\star}(T_\text{FO}) }{g^{\omega}_{\star}(T_{\nu\text{dec}})}\dfrac{g^{\text{SM}}_{\star}(T_{\nu\text{dec}})}{g^{\text{SM}}_{\star}(T_\text{FO})}~.
\end{equation}
At a photon temperature $T\ll T_{\nu\text{dec}}$ by entropy conservation we have
\begin{align}
            g^{\omega}_{\star}(T)T_\omega^3(T) &=g^{\omega}_{\star}(T_{\nu\text{dec}})T_\omega^3(T_{\nu\text{dec}})~, \nonumber \\
            g^{\nu}_{\star}(T)T_\nu^3(T)&=g^{\nu}_{\star}(T_{\nu\text{dec}})T_{\nu\text{dec}}^3~, \\
g^{\gamma}_{\star}(T)T^3 &=g^{\gamma}_{\star}(T_{\nu\text{dec}})T_{\nu\text{dec}}^3~, \nonumber
\end{align}
which give the following expression for the neutrino temperature
\begin{equation}
    T_\nu(T)=\left( \dfrac{g^{\gamma}_{\star}(T)}{g^{\gamma}_{\star}(T_{\nu\text{dec}})}\right)^{1/3} T~,
\end{equation}
with $g^{\gamma}_{\star}(T)=2$ and $g^{\gamma}_{\star}(T_{\nu\text{dec}})=2+(7/8)\times 4 =11/2$ we recover the usual SM relation
\begin{equation}
    T_\nu(T)=\left( \dfrac{4}{11}\right)^{1/3} T~.
\end{equation}
For the dark sector we have
\begin{equation}
        \dfrac{T_\omega^3(T)}{T^3}=\dfrac{g^{\gamma}_{\star}(T)}{ g^{\gamma}_{\star}(T_{\nu\text{dec}}) }
        \dfrac{g^{\omega}_{\star}(T_{\nu\text{dec}})}{g^{\omega}_{\star}(T)}\dfrac{T_\omega^3(T_{\nu\text{dec}})}{T_{\nu\text{dec}}^3}= \dfrac{g^{\gamma}_{\star}(T)}{ g^{\gamma}_{\star}(T_{\nu\text{dec}}) }
        \dfrac{g^{\omega}_{\star}(T_{\text{FO}})}{g^{\omega}_{\star}(T)}
\dfrac{g^{\text{SM}}_{\star}(T_{\nu\text{dec}})}{g^{\text{SM}}_{\star}(T_\text{FO})}~.
\end{equation}
\begin{center}
\begin{table}[t!]
    \centering
   \begin{tabular}{ c || c | c | c }
   \text{Relativistic species at} $T_\text{FO}$ &  $g_\star^\omega(T_\text{FO})$ & $\Delta N_\text{eff} ~(T_\text{FO}=T_\text{QCD})$ &$\Delta N_\text{eff}~(T_\text{FO}>T_\text{EW})$\\
    \hline
    $\omega$ &  2 & 0.11 & 0.047    \\
    \hline
    $\omega+\chi$ & 6 & 0.49 & 0.2   \\
    \hline
    $\omega+N_{4,5}$ & 14 & 1.51 & 0.62   \\
    \hline
    $\omega+\chi+Z^\prime$ & 66/7 & 0.89 & 0.37   \\
    \hline
    $\omega+\chi+Z^\prime+N_{4,5}$ & 150/7 & 2.67 & 1.11  \\
  \end{tabular}
      \caption{Effective relativistic fermionic degrees of freedom of the dark sector for a given specific particle content at $T_\text{FO}$ and corresponding contributions to $\Delta N_\text{eff}$, assuming the freeze-out to occur at the QCD phase transition $T_\text{FO}=T_\text{QCD}$ and above the electroweak phase transition $T_\text{FO}>T_\text{EW}$}
    \label{tab:DeltaNeff}
\end{table}
\end{center}
At low temperature $T$, the ratio of the dark sector to photon energy density is given by
\begin{equation}
    \dfrac{\rho_\omega(T)}{\rho_\gamma(T)}=\dfrac{7}{8}\dfrac{g_{\star}^\omega(T) T_{\omega}^4(T)}{g_{\star}^\gamma(T)T^4}~.
\end{equation}
Taking $g_{\star}^\omega(T)=g_{\star}^\gamma(T)=2$ for $T\ll T_{\nu \text{dec}}$, $\Delta N_\text{eff} $ is given by 
\begin{equation}
    \Delta N_\text{eff} =  \left( \dfrac{11}{4} \right)^{4/3} \dfrac{g_{\star}^\omega(T) T_{\omega}^4(T)}{g_{\star}^\gamma(T)T^4}=  \left( \dfrac{11}{4}   \dfrac{g^{\omega}_{\star}(T_{\text{FO}})}{ g^{\gamma}_{\star}(T_{\nu\text{dec}})}
\dfrac{g^{\text{SM}}_{\star}(T_{\nu\text{dec}})}{g^{\text{SM}}_{\star}(T_\text{FO})}   \right)^{4/3}~.
\end{equation}
At the neutrino decoupling we have $g^{\text{SM}}_{\star}(T_{\nu\text{dec}})=2+(7/8)\times( 4+2\times3)=43/4$ and $g^{\gamma}_{\star}(T_{\nu\text{dec}})=2+(7/8)\times 4 =11/2$, giving the following expression for $\Delta N_\text{eff} $:
\begin{equation}
    \Delta N_\text{eff} =\left( \dfrac{43}{4} \right)^{4/3} \left( \dfrac{g^{\omega}_{\star}(T_{\text{FO}})}{2} \right)^{4/3}  \left(  \dfrac{1}{g^{\text{SM}}_{\star}(T_\text{FO})}   \right)^{4/3}~.
\end{equation}

For illustration, in Tab.~\ref{tab:DeltaNeff} we reported the expected values for $\Delta N_\text{eff}$ given a specific relativistic content at the freeze-out temperature. We estimate the effective degrees of freedom of the SM using the approximate fitted expression of Appendix A of Ref.~\cite{Wantz:2009it}.
One can deduce from Tab.~\ref{tab:DeltaNeff}, given the large expected values of $\Delta N_\text{eff}$, that most of the states of the dark sector must become non-relativisitic before the freeze-out temperature. In particular, contributions from the heavy neutrinos $N_{4,5}$ to $\Delta N_\text{eff}$ would exceed 0.5 if these states are still relativistic at the freeze-out time due to their large multiplicity, unless some mass hierarchy is present among them.

\section{Collision terms}
\label{sec:appendix_collision_terms}

\subsection*{$Z^\prime$ inverse-decay}

The Boltzmann equation corresponding to $Z^\prime$ production via $\bar\psi(p_1) + \psi(p_2) \leftrightarrow Z^\prime (p_3)$ is given by Eq.~(\ref{eq:Boltzmann_inversdecay}) whose right-hand-side can be written in terms of the $Z^\prime$ yield $Y_{Z^\prime}\equiv n_{Z^\prime}/s$ as:
\begin{equation}
    \begin{split}
H(z)s(s)z\frac{\diff Y_{Z^\prime}}{\diff z}  =  \int \prod_{i=1}^3 \frac{\diff^3 \vec p_i}{(2\pi)^3 2 E_i} & \left[    |{\cal A}_{\psi\bar \psi \to Z^\prime}|^2 f_1(\vec p_1) f_2(\vec p_2) \right. \\ & \left.  - |{\cal A}_{Z^\prime \to \psi\bar \psi}|^2 f_3(\vec p_3)   \right] (2\pi)^4 \delta^4(p_1+p_2 - p_3)~,       
    \end{split}
\end{equation}
 where  $z \equiv m_{Z^\prime}/T$ and we neglected temperature dependence of the relitivistic degrees of freedom. $s$ and $H$ are respectively the entropy density and Hubble expansion rate.
By detailed balance, we can simplify the inverse process using 
\be
 |{\cal A}_{\psi \bar \psi \to Z^\prime}|^2 f^{\rm (eq)}_1(\vec p_1) f^{\rm (eq)}_2(\vec p_2)   =  |{\cal A}_{Z^\prime \to \psi \bar \psi}|^2f^{\rm (eq)}_3(\vec p_3)~,
\ee
so the Boltzmann equation becomes
\be \label{eq:boltzt}
H(z)s(s)z\frac{\diff Y_{Z^\prime}}{\diff z}   =
 \int   \frac{\diff^3 \vec p_3}{(2\pi)^3 }  \dfrac{m_{Z^\prime}}{E_3}  \Gamma_{Z^\prime \to \bar \psi \psi}(E_3)    \left[  f^{\rm (eq)}_3(\vec p_3)  -     
 f_3(\vec p_3)             \right] = \langle     \Gamma_{Z^\prime \to \bar \psi \psi}     \rangle   \left[  n_{Z^\prime}^{\rm (eq)}(t)  -    n_{Z^\prime}(t) \right],
 \ee
where we have defined 
\be
\langle  \Gamma_{Z^\prime \to \bar \psi \psi}\rangle \equiv \frac{\int  \diff^3 \vec p_3  \,     \Gamma_{Z^\prime \to \bar \psi \psi} (m_{Z^\prime}/E_3)  e^{-E_3/T}    }{
\int   \diff^3 \vec p_3   \,  e^{-E_3/T}} 
=  \Gamma_{Z^\prime \to \bar \psi \psi} \frac{ K_1(z)  }{  K_2( z) } ~,~~
\ee
using Maxwell-Boltzmann distributions. The Boltzmann equation can therefore be written as 
\be 
 \frac{\diff Y_{Z^\prime}}{\diff z}  =  \frac{    \Gamma_{Z^\prime \to \bar \psi \psi}  }{ H(z)  z  } \frac{ K_1(z) }{  K_2(z) }   \left[  Y_{Z^\prime}^{\rm (eq)}(z)  -    Y_{Z^\prime}(z) \right],
 \ee
 where $K_{1,2}$ are modified Bessel functions of the second kind and we used the $Z^\prime$ expected equilibrium yield $Y_{Z^\prime}^{\rm (eq)}(z)$ given by
\begin{equation}
 Y_{Z^\prime}^{\rm (eq)}(z)  = \frac{3 m_{Z^\prime}^3}{2 \pi^2 s(z)z^3}       \int_{0 }^\infty     \dfrac{    1 }{ e^{\sqrt{y^2+z^2}}    -1 }  y^2 \diff y~,
\end{equation}
where $y\equiv p/T$ with $p$ being the $Z^\prime$ momentum.

\subsection*{$2\rightarrow 2$ annihilations}
Collision terms $\mathcal{C}$ corresponding to production of a given state in $2\rightarrow 2$ processes considered in this work are parametrized by the following term
\begin{equation}
\mathcal{C}=  \int \prod_{i=1}^4 \frac{\diff^3 \vec p_i}{(2\pi)^3 2 E_i} (2\pi)^4 E_1^\alpha E_2^\beta f_1(\vec p_1) f_2(\vec p_2) |{\cal A}_{1 2 \to 3 4}|^2 \delta^4(p_1+p_2- p_3 -p_4 )~,
\end{equation}
where the produced state is either $3$ or $4$, or both, and the backreaction term is neglected. $E_i=\sqrt{p_i^2+m_i^2} $ is the energy of the state $i$ with momentum $p_i$ and corresponding phase space distribution $f_i(\vec{p}_i)$. $|{\cal A}_{1 2 \to 3 4}|^2 $ is the corresponding matrix element squared for this process summed over initial and final-states internal degrees of freedom and containing symmetry factors accounting for identical initial or final states. $\alpha$ and $\beta$ are integers. In the limit where all state are relativistic, collision terms can be written as
\begin{equation}
\mathcal{C}=  \int \dfrac{E_1^{1+\alpha}E_2^{1+\beta}\diff E_1 \diff E_2 \diff \cos \theta_{12}}{1024 \pi^6} \int |{\cal A}_{1 2 \to 3 4}|^2 \diff \Omega_{13}~,
\end{equation}
where $\theta_{12}$ is the angle formed by the initial-states momenta and $\Omega_{13}$ the solid angle betweem particle $1$ and $3$. Assuming that the amplitude grows as powers of $|{\cal A}_{1 2 \to 3 4}|^2 = s^n/\Lambda^{2n}$ where $s=2p_1 p_2(1-\cos \theta_{12})$ is the typical Mandelstam variable with $n$ being an integer and $\Lambda$ an energy scale, analytical expressions can be derived for collision terms:  

\begin{equation}
  \mathcal{C}=   \frac{\Gamma (n+\alpha +2) \Gamma (n+\beta +2) \text{Li}_{n+\alpha +2}(-\epsilon_1) \text{Li}_{n+\beta +2}(-\epsilon_2) }{\pi ^5 (n+1) \epsilon_1 \epsilon_2 2^{7-2 n} \Lambda ^{2 n}} T^{\alpha +\beta +2 n+4}~,
  \label{eq:collisionterms_analytical}
\end{equation}
for phase space distributions of initial state particles $f(p_i)=1/(\exp(p_i/T)+\epsilon_i 1)$ corresponding to Fermi-Dirac statistics for $\epsilon_i=1$ and Bose-Einstein for $\epsilon_i=-1$. $\text{Li}_{n}(x)$ are polylogarithm special functions. In the limit of Maxwell-Boltzmann statistics, the previous reduces to  
\begin{equation}
\mathcal{C}=  \frac{ \Gamma (n+\alpha +1) \Gamma (n+\beta +1) }{\pi ^5 (n+1) 2^{7-2 n} \Lambda ^{2 n}}T^{\alpha +\beta +2 n+4}~,
\end{equation}
which shows discrepancies with Fermi-Dirac or Bose-Einstein statistics by a factor of $\sim 3$ at most, depending on the values of the integers $\alpha, \beta, n$. In the following we present analytical expressions of collision terms used in this work based on Eq.~(\ref{eq:collisionterms_analytical}). In some places, we used the symbol $R$ to denote dimensionless reaction rates with the general definition:
\begin{equation}
    R(z)\equiv \dfrac{1}{H(z)s(z)z} \dfrac{\delta n}{\delta t} ~.
\end{equation}

\subsubsection*{$e^\pm \gamma \leftrightarrow e^\pm Z^\prime$}
The collision term defined in Eq.~(\ref{eq:collision_term_egamma_to_eZprime}) is given by
\begin{equation}
   \left. \dfrac{\delta n_{Z^\prime}}{\delta t} \right|_{e^\pm \gamma \to e^\pm Z^\prime}=\frac{e^4 \varepsilon^2 T^4}{288 \pi }~.
\end{equation}

\subsubsection*{$e^+ e^- \rightarrow \omega \omega$}
The collision term used to compute the rate of Eq.~(\ref{eq:rate2to2omega}) is given by
\begin{equation}
   \left. \dfrac{\delta n_\omega}{\delta t} \right|_{e^+ e^- \to \omega \omega} =\frac{e^2 g_{Z^\prime}^2 \varepsilon^2 T^4}{288 \pi }~.
\end{equation}
\subsubsection*{$\bar \psi \psi \rightarrow \omega \omega$}
The collision term for the energy-density Boltzmann equation defined in
Eq.~(\ref{eq:collision_term_psipsitoomegaomega}) is is given by
\begin{equation}
 \left.\dfrac{\delta \rho_\omega}{\delta t}\right|_{\bar \psi \psi \to \omega \omega} = c_\psi Q_\psi^2 \dfrac{7\zeta(5)}{2 \pi}\dfrac{  e^2 \varepsilon^2 g_{Z^{\prime}}^2 T^9}{ m_{Z^\prime}^4}~.
\end{equation}

\subsubsection*{$\bar t t \rightarrow \bar \chi \chi$}
The collision term used to compute the rate of Eq.~(\ref{eq:collision_term_ttbartochichi}) is given by
\begin{equation}
   \left. \dfrac{\delta n_\chi}{\delta t} \right|_{\bar t t \to \bar \chi \chi} =\sin^2 \alpha \frac{y_{\chi }^2 m_t^2  T^4}{3072 \pi  v_h^2}~.
\end{equation}

\end{document}